\documentclass[physrev,aps,twocolumn,amsmath,amssymb,superscriptaddress]{revtex4-2}
\pdfoutput=1

\usepackage[latin9]{inputenc}
\usepackage{epsfig}
\usepackage{times} 
\usepackage{ucs}
\usepackage{bbm}
\usepackage{orcidlink} 

\usepackage{hyperref}
\hypersetup{colorlinks=true, citecolor=blue, urlcolor=blue, linkcolor=blue}

\newcommand{\bra}[1]{\langle #1|}
\newcommand{\ket}[1]{|#1\rangle}
\newcommand{\ketbra}[1]{| #1\rangle \langle #1|}

\newcommand{\be}{\begin{equation}}
\newcommand{\ee}{\end{equation}}
\newcommand{\eea}{\end{eqnarray}}
\newcommand{\bea}{\begin{eqnarray}}

\newcommand{\va}[1]{\ensuremath{(\Delta#1)^2}}
\newcommand{\vasq}[1]{\ensuremath{[\Delta#1]^2}}

\newcommand{\ex}[1]{\ensuremath{\left\langle{#1}\right\rangle}}
\newcommand{\exs}[1]{\ensuremath{\langle{#1}\rangle}}

\newcommand{\qed}{\ensuremath{\hfill \blacksquare}}

\newcommand{\kommentar}[1]{}
\newcommand{\trace}{{\rm Tr}}

\newcommand{\forget}[1]{}

\newcommand{\EQ}[1]{Eq.~\eqref{#1}}
\newcommand{\EQS}[1]{Eqs.~\eqref{#1}}
\newcommand{\EQL}[1]{Equation~\eqref{#1}}
\newcommand{\SEC}[1]{Sec.~\ref{#1}}
\newcommand{\FIG}[1]{Fig.~\ref{#1}}
\newcommand{\REF}[1]{Ref.~\cite{#1}}
\newcommand{\REFS}[1]{Refs.~\cite{#1}}
\newcommand{\REFL}[1]{Reference~\cite{#1}}

\newcommand{\APP}[1]{Appendix~\ref{#1}}
\newcommand{\OBS}[1]{Observation~\ref{#1}}

\usepackage{ntheorem}
\theoremheaderfont{\hspace*{\parindent}\bf}
\theoremseparator{.}
\theorempreskip{0cm}
\theorempostskip{0cm}
\newtheorem{observation}{Observation}
\let\oldobservation\observation
\renewcommand{\observation}{\oldobservation\normalfont}

\begin{document}

\title{Uncertainty relations with the variance and the quantum Fisher information based on convex decompositions of density matrices}

\author{G\'eza T\'oth\,\orcidlink{0000-0002-9602-751X}}
\email{toth@alumni.nd.edu}
\homepage{http://www.gtoth.eu}
\affiliation{Department of Theoretical Physics, University of the Basque Country
UPV/EHU, P.O. Box 644, E-48080 Bilbao, Spain}
\affiliation{Donostia International Physics Center (DIPC), P. O. Box 1072, E-20080 San Sebasti\'an, Spain}
\affiliation{IKERBASQUE, Basque Foundation for Science, E-48013 Bilbao, Spain}
\affiliation{Institute for Solid State Physics and Optics, Wigner Research Centre for Physics,
P. O. Box 49, H-1525 Budapest, Hungary}
\author{Florian Fr\"owis\,\orcidlink{0000-0002-2743-3119}}
\affiliation{Group of Applied Physics, University of Geneva, CH-1211 Geneva, Switzerland}

\begin{abstract}
We present several inequalities related to the Robertson-Schr\"odinger uncertainty relation. In all these inequalities, we consider a decomposition of the density matrix into a mixture of states, and use the fact that the Robertson-Schr\"odinger uncertainty relation is valid for all these components. By considering a convex roof of the bound, we obtain an alternative derivation of the relation in Fr\"owis {\it et al.} \href{https://doi.org/10.1103/PhysRevA.92.012102}{[Phys. Rev. A {\bf 92}, 012102 (2015)]}, and we can also list a number of conditions that  are needed to saturate the relation. We present a formulation of the Cram\'er-Rao bound involving the convex roof of the variance. By considering a concave roof of the bound in the Robertson-Schr\"odinger uncertainty relation over decompositions to mixed states, we obtain an improvement of the Robertson-Schr\"odinger uncertainty relation. We consider similar techniques for uncertainty relations with three variances. Finally, we present further uncertainty relations that provide  lower bounds on the metrological usefulness of bipartite quantum states based on the variances of the canonical position and momentum operators for two-mode continuous variable systems. We show that the violation of well-known entanglement conditions in these systems discussed in Duan {\it et al.}, \href{https://doi.org/10.1103/PhysRevLett.84.2722}{[Phys. Rev. Lett. {\bf 84}, 2722 (2000)]} and Simon \href{https://doi.org/10.1103/PhysRevLett.84.2726}{[Phys. Rev. Lett. {\bf 84}, 2726 (2000)]} implies that the state is more useful metrologically than certain relevant subsets of separable states. We present similar results concerning entanglement conditions with angular momentum operators for spin systems.

\vspace{1em}
\noindent DOI: \href{https://doi.org/10.1103/PhysRevResearch.4.013075}{10.1103/PhysRevResearch.4.013075}
\end{abstract}

\date{\today}

\maketitle

\section{Introduction}

Quantum Fisher information (QFI) is a central quantity of quantum metrology, a field that is concerned with metrological tasks in which the quantumness of the system plays an essential role  \cite{Giovannetti2004Quantum-Enhanced,Paris2009QUANTUM,Demkowicz-Dobrzanski2014Quantum,Pezze2014Quantum}. One of the most fundamental scenarios in quantum metrology is estimating the small parameter $\theta$ in the unitary dynamics 
\begin{equation}
\varrho_\theta=e^{-iB\theta}\varrho e^{iB\theta},
\label{eq:rho_theta}
\end{equation}
where $B$ is the Hamiltonian of the dynamics, $\varrho$ is the initial state, $\varrho_\theta$ is the final state of the evolution, and we set $\hbar=1$ for simplicity. By carrying out measurements on $\varrho_{\theta},$ we aim to estimate $\theta$ from the distribution of the outcomes of the measurement. The quantum Cram\'er-Rao inequality gives a lower bound on the precision of the estimation for any measurement 
\begin{equation} \label{eq:cra}
\va{\theta}\ge \frac{1}{m F_Q[\varrho, B]},
\end{equation}
where $F_Q[\varrho, B]$ is the QFI and $m$ is the number of independent repetitions \cite{Helstrom1976Quantum,Holevo1982Probabilistic,Braunstein1994Statistical, Braunstein1996Generalized}. 

At the center of attention lies the question how noise can affect the precision of the estimation \cite{Huelga1997Improvement} and what the ultimate limit of the precision is in realistic scenarios \cite{Escher2011General,Demkowicz-Dobrzanski2012The}. We add that a driving force behind the development in quantum metrology are recent experiments in quantum optical systems, such as cold gases and cold trapped ions, which are possible due to the rapid technological advancement in the field \cite{Leibfried2004Toward,Napolitano2011Interaction-based,Riedel2010Atom-chip-based,Gross2010Nonlinear}. The experiments with the squeezed-light-enhanced gravitational wave detector GEO 600 \cite{Grote2013First,Aasi2013Enhanced,Demkowicz-Dobrzanski2013Fundamental} are highlights in the applications of quantum-enhanced sensitivity.

Recently, the QFI was discovered to play an important role in quantum information theory, in particular, in the theory of quantum entanglement \cite{Toth2014Quantum}. It turns out that, in linear interferometers,  entanglement is needed to surpass the shot-noise limit in precision corresponding to product states \cite{Giovannetti2004Quantum-Enhanced,Pezze2009Entanglement,Paris2009QUANTUM,Demkowicz-Dobrzanski2014Quantum,Pezze2014Quantum}. It has been shown that the larger the QFI, the larger the depth of entanglement  the state must posses \cite{Hyllus2012Fisher,Toth2012Multipartite}. Beside the  entanglement depth, there are further quantities that can give a more detailed information about the structure of the multipartite entanglement \cite{Szalay2019Stretchability,Toth2020Stretching}, which turn out to be strongly connected to the QFI \cite{Ren2021Metrological}. In general, the QFI can be used to detect multipartite entanglement, which has been done in several experiments \cite{Lucke2011Twin,Krischek2011Useful,Strobel2014Fisher}. Apart from entanglement theory, the QFI has also been used to define what it means that a superposition is macroscopically quantum \cite{Frowis2011Stable,Frowis2012Are}, and to bound the speed of a quantum evolution \cite{Campo2013Quantum,Taddei2013Quantum}, and it plays a role even in the quantum Zeno effect \cite{Smerzi2012Zeno,Schafer2014Experimental}. Finally, the QFI offers a powerful characterization of the prepared quantum state, for which it is calculated even from tomographic data \cite{Schwemmer2014Experimental}. It has been shown that this type of characterization is superior to computing the fidelity with respect to the ideal state, for usual state reconstruction schemes \cite{Schwemmer2015Systematic}.

Recent findings show that the QFI is the convex roof of the variance, apart from a constant factor \cite{Toth2013Extremal,Yu2013Quantum}.  This again connects quantum metrology to quantum information science where convex roofs often appear in the theory of entanglement measures \cite{Horodecki2009Quantum,Guhne2009Entanglement}. Density matrices have an infinite number of convex decompositions. This is a feature of quantum mechanics, not present in classical physics. So far, this fact is appreciated mostly in quantum information science, however, it can also be used as a powerful tool in  other areas of quantum physics.

Finally, the QFI appears in various quantum uncertainty relations. In these relations, the error propagation formula defined as 
\begin{equation}\label{eq:errprop}
(\Delta \theta)^2_A  = \frac{(\Delta A)^2}{\left| \partial_{\theta} \exs{A}  \right|^2}
\end{equation}
plays a central role \cite{Frowis2015Tighter}. The uncertainty of the estimate is given by \EQ{eq:errprop} divided by $m,$ the number of independent repetitions, if the distribution of the measurement results fulfill certain reasonable requirements and $m$ is sufficiently large \cite{Kholevo_Generalization_1974,Braunstein1994Statistical,Zhong_Optimal_2014}. Then, from the Cram\'er-Rao bound (\ref{eq:cra}), one can derive \cite{Frowis2015Tighter}, for example, the Heisenberg-Robertson uncertainty relation \cite{Heisenberg1927Uber,Robertson1929The}, time-energy uncertainty relations \cite{Mandelstam1945TheB,Fleming_unitarity_1973,Uhlmann_energy_1992,Frowis2012Kind} and squeezing inequalities \cite{Sorensen2001Many-particle,Pezze2009Entanglement}. The optimization of \EQ{eq:errprop} over a given set of operators has been considered \cite{Gessner2019Metrological}.

In this paper, we use the knowledge that the QFI is, apart from a constant factor, the convex roof of the variance to obtain inequalities valid for all quantum states, and to obtain entanglement criteria. First, we give a simple proof of a tighter version of the Heisenberg-Robertson uncertainty relation  \cite{Schrodinger1930Zum, Frowis2015Tighter}, also giving conditions for saturation.   We show ways to strengthen the Heisenberg-Robertson uncertainty relation. We derive the Cram\'er-Rao bound such that the bound is given by a convex roof. We derive a relation with two variances and a QFI. We also present entanglement conditions with the QFI.

Our paper is organized as follows.
In \SEC{sec:Important_properties}, we summarize important properties of the QFI and the variance. 
In \SEC{sec:conn-betw-recent}, we discuss recent finding connecting the QFI to convex roofs. 
In \SEC{sec:convroof}, we present inequalities derived from the Robertson-Schr\"odinger uncertainty relation based on convex roofs.
In \SEC{sec:RS}, we present an improvement on the same inequality based on concave roofs.
In \SEC{sec:several_var}, we present uncertainty relations with variances and the QFI.
In \SEC{sec:alt}, we present simple relation and use it to rederive some of our results. We also derive further inequalities with the variance and the QFI.
In \SEC{sec:var_metrlogy}, we show how to relate the violation of  some entanglement conditions to the metrological usefulness of the quantum state.

\section{Important properties of the QFI}
\label{sec:Important_properties}

In this section we briefly summarize the basic literature about the QFI. The properties we list will be used later in our calculations.

Most importantly, the QFI is convex, i.e.,
\begin{equation}\label{eq:convexity}
F_Q[\varrho_{\rm m},B]\le pF_Q[\varrho_1,B]+(1-p)F_Q[\varrho_2,B],
\end{equation}
where the mixture is defined as
\begin{equation}
\varrho_{\rm m}=p\varrho_1+(1-p)\varrho_2.
\end{equation}
Here lies an important similarity between the QFI and entanglement measures: neither of the two can increase under mixing.

The QFI appearing in the Cram\'er-Rao bound \EQ{eq:cra} is defined as \cite{Helstrom1976Quantum,Holevo1982Probabilistic,Braunstein1994Statistical,
Petz2008Quantum,Braunstein1996Generalized}
\begin{equation}
F_{Q}[\varrho,A]=2\sum_{k,l}\frac{(\lambda_{k}-\lambda_{l})^{2}}{\lambda_{k}+\lambda_{l}}\vert \bra{k}A\ket{l}\vert^{2},\label{eq:qF}
\end{equation}
where the density matrix has the eigendecomposition
\begin{equation}\label{eq:rho_eigdecomp}
\varrho=\sum_{k}\lambda_k \ketbra{k}.
\end{equation}
From \EQ{eq:qF}, it follows that the QFI can be bounded from above by the variance 
\begin{equation}
F_Q[\varrho,B] \leq 4(\Delta B)^2_{\varrho}\label{eq:FQvar},
\end{equation}
where equality holds if $\varrho$ is pure \cite{Braunstein1994Statistical}.

The Cram\'er-Rao bound \eqref{eq:cra} defines the achievable largest precision of parameter estimation, however, it is not clear what has to be measured to reach this precision bound. An optimal measurement can be carried out if we measure in the eigenbasis of the symmetric logarithmic derivative $\mathcal{L}$  \cite{Braunstein1994Statistical,Braunstein1996Generalized}. This operator is defined such that it can be used to describe the quantum dynamics of the system with the equation
\begin{equation}\label{eq:LrrL}
\frac{d\varrho_\theta}{d\theta}=\tfrac{1}{2}(\mathcal{L}\varrho_\theta+\varrho_\theta \mathcal{L}).
\end{equation}
Unitary dynamics are generally given by the von Neumann equation with the Hamiltonian $B$
\begin{equation}\label{eq:LrrL2}
\frac{d\varrho_\theta}{d\theta}=i(\varrho_\theta B-B\varrho_\theta).
\end{equation}
The operator $\mathcal{L}$ can be found based on knowing that the right-hand side of \EQ{eq:LrrL} must be equal to the right-hand side of \EQ{eq:LrrL2}:
\begin{equation}
i[\varrho,B] = \tfrac{1}{2} \left\{ \varrho,\mathcal{L}\right\}.\label{eq:comm_anticomm2}
\end{equation}
Hence, the symmetric logarithmic derivative can be expressed with a simple formula as 
\begin{equation}
\label{eq:L}
\mathcal{L}=2i\sum_{k,l}\frac{\lambda_{k}-\lambda_{l}}{\lambda_{k}+\lambda_{l}} \vert k \rangle \langle l \vert \langle k \vert B \vert l \rangle,
\end{equation}
where $\lambda_k$ and $\vert k\rangle$ are the eigenvalues and eigenvectors, respectively, of the density matrix $\varrho.$ Based on \EQS{eq:qF} and \eqref{eq:L}, the symmetric  logarithmic derivative  can be used to obtain the QFI as
\begin{equation}
F_Q[\varrho,B]=\trace(\varrho \mathcal{L}^2)=(\Delta \mathcal{L})^2.\label{eq:LLL}
\end{equation}
In the second equality in \EQ{eq:LLL}, we used that 
\begin{equation}
\exs{\mathcal{L}}_{\varrho}=0,\label{eq:L0}
\end{equation}
which can be seen based on \EQS{eq:rho_eigdecomp} and  \eqref{eq:L}.

\section{Defining the quantum Fisher information with convex roofs}
\label{sec:conn-betw-recent}

The quantum Fisher information has been connected to convex roofs that are based on an optimization over convex decompositions of the density matrix \cite{Toth2013Extremal,Yu2013Quantum}. Let us consider a density matrix of the form
\begin{equation}
\varrho=\sum_k p_k \ketbra{\psi_k}, \label{decomp}
\end{equation}
where $p_k>0$ and $\sum_k p_k=1.$ 
Note that the pure states $\ket{\psi_{k}}$ are not required to be pairwise orthogonal, and \EQ{decomp} is not an eigendecomposition of the density matrix. Then, it can be shown that the QFI is the convex roof of the variance times four \cite{Toth2013Extremal,Yu2013Quantum}
\begin{equation}
F_{Q}[\varrho,B] =4 \min_{\{ p_k,\ket{\psi_k}\}} \sum_k p_k \va {B
 }_{\psi_k},\label{e2b}
\end{equation}
where $\{p_{k},\ket{\psi_{k}}\}$ refers to a decomposition of $\varrho$ of the type Eq.~(\ref{decomp}). In other words, we already knew that the QFI is convex, but \EQ{e2b} implies that it is the smallest convex function that equals four times the variance for pure states. For further analysis on the convexity of the QFI, see \REF{Rezakhani2019Continuity}.

\EQL{e2b} has also been used in derivations concerning the continuity of the QFI \cite{Augusiak2016Asymptotic}, or finding efficient ways to bound it from below based on few measurements \cite{Apellaniz2017Optimal}. It has been used in constructing entanglement conditions in \REF{Akbari-Kourbolagh2019Entanglement}. Finally, it has also been used in finding a bound on 
\begin{equation}
V(\varrho,A)=\va{A}-F_Q[\varrho,A]/4\label{eq:vaminusFQ}
\end{equation}
based on the purity of $\varrho$ \cite{Toth2017Lower}.

A related result is that the variance is the concave roof of itself
\begin{equation}\label{eq:Fisherroof}
\va A_{\varrho}=\max_{\{ p_k,\ket{\psi_k}\}} \sum_k p_k \va A_{\psi_k}. 
\end{equation}
This property of the variance is relatively easy to show \cite{Toth2013Extremal,Yu2013Quantum}. For the proof, one has to demonstrate that there is always a decomposition of the type \EQ{decomp} such that 
\begin{equation}
\exs{A}_{\psi_k}=\exs{A}_{\varrho}\label{eq:exApsi}
\end{equation}
for all $k.$ Similar decompositions for correlation matrices have been considered in  Refs.~\cite{Leka2013Some,Petz2014}.

The statements of \EQS{e2b} and \eqref{eq:Fisherroof} can be concisely reformulated as follows. For any decomposition $\{p_{k},\ket{\psi_{k}}\}$ of the density matrix $\varrho$ we have
\begin{equation}
\frac{1}{4}F_{Q}[\varrho,A]  \le  \sum_k p_k \va {A}_{\psi_k} \le \va A_{\varrho},\label{eq:FsumV}
\end{equation}
where the upper and the lower bounds are both tight. 

Note that the QFI has been connected to convex roofs in another context, via purifications  \cite{Fujiwara2008A,Escher2011General,Demkowicz-Dobrzanski2012The,Marvian2020Coherence}. The basic idea is that the QFI can easily be computed for pure states and a unitary dynamics. For the more general case of mixed states and noisy dynamics we can still deal with pure states, if we add an ancillary system and consider the purification of the noisy dynamics.

Finally, let us discuss that relations given in \EQS{e2b} and \eqref{eq:Fisherroof} remain the same if we optimize over decompositions to a mixture of density matrices instead of decompositions to a mixture of pure states. Let us consider a decomposition of $\varrho$ to a mixture of density matrices \cite{*[{Decomposition to mixed states also appears in entanglement theory, e.g., see  }] [{.}]  Hofmann2014Analytical}
\begin{equation}
\varrho=\sum_k p_k \varrho_k.\label{eq:rhodecomp}
\end{equation}
Due to the fact that the QFI and the variance are convex and concave, respectively, in density matrices, the inequalities
\begin{equation}
\frac{1}{4}F_{Q}[\varrho,A]  \le  \sum_k p_k \va {A}_{\varrho_k} \le \va A_{\varrho}\label{eq:FsumV2}
\end{equation}
hold. However, we already know that decompositions to a mixture of pure states can saturate both inequalities in \EQ{eq:FsumV}. Thus, obtaining the convex and concave roofs over decompositions to mixed states will lead to same values that we obtain in \EQS{e2b} and \eqref{eq:Fisherroof}.

Concerning the relation for the QFI given in \EQ{e2b}, we can add the following. If we calculate the convex roof of a quantity that is concave in density matrices, then the result of a minimization over all pure-state decompositions will coincide with the result of a minimization over all mixed-state decompositions. The reason is that if a concave function is minimized over a convex set, then it takes its minima on the extreme points of the set. Similarly, if we calculate the concave roof of a quantity that is convex in density matrices, then the result of a maximization over all pure-state decompositions will coincide with the result of a maximization over all mixed-state decompositions.

\section{Uncertainty relations  based on a convex roof over decompositions in the Robertson-Schr\"odinger inequality}
\label{sec:convroof}

The Robertson-Schr\"odinger uncertainty is a fundamentally important uncertainty relation in quantum physics \cite{Robertson1929The}. Hence,  there is a strong interest in deriving further relations from it and in looking for possible improvements \cite{Luo2005Heisenberg,Furuichi2010Schrodinger,Yu2013Robertson,Maccone2014Stronger}. In this section, we present a simple method to obtain further uncertainty relations based on the optimization over the decompositions of density matrices. We rederive the improved Heisenberg-Robertson inequality presented in \REF{Frowis2015Tighter}. We discuss some implications of the Cram\'er-Rao bound, and determine, which states saturate the inequality.
 
\subsection{Simple proof for the improved Heisenberg-Robertson inequality presented in \REF{Frowis2015Tighter}}
\label{subsec:Simple_proof}

The Robertson-Schr\"odinger inequality is defined as 
\begin{equation}
\label{eq:RS1}
\va{A}_\varrho \va{B}_\varrho
\ge \tfrac{1}{4} \vert L_\varrho \vert^2,
\end{equation}
where the lower bound is given by 
\begin{eqnarray}\label{eq:defL}
L_\varrho=\sqrt{{\vert \exs{\{A,B\}}_\varrho-2\exs{A}_\varrho\exs{B}_\varrho\vert^2+\vert \exs{C}_\varrho\vert^2}},
\end{eqnarray}
$\{A,B\}=AB+BA$ is the anticommutator, and we used the definition 
\begin{equation}
C=i[A,B].\label{eq:CAB}
\end{equation}

First let us examine the convexity properties of  the bound on the right-hand side of \EQ{eq:RS1}, which we need later. One can show that $L_\varrho$ is neither convex nor concave in $\varrho.$  Let us consider a concrete example, the mixed two-qubit state with a decomposition 
\begin{align}
p_1&=1/2, &\ket{\psi_1}&=\ket{00}, \nonumber\\
p_2&=1/2, &\ket{\psi_2}&=\ket{11}, 
\end{align}
and the operators 
\begin{eqnarray}
A&=\sigma_z \otimes \openone, \nonumber\\
B&= \openone \otimes \sigma_z. \label{eq:AB2}
\end{eqnarray}
For these, we have $C=0$, $L_{\psi_1}=L_{\psi_2}=0,$ while $L_{\varrho}=1.$ Simple algebra shows that $L_{\varrho}>p_1L_{\psi_1}+p_2L_{\psi_2}$ holds. Let us consider another concrete example, the mixed state with a decomposition 
\begin{align}
p_1&=1/2, &\ket{\psi_1}&=(\ket{00}+\ket{11})/\sqrt{2}, \nonumber\\
p_2&=1/2, &\ket{\psi_2}&=(\ket{01}+\ket{10})/\sqrt{2}, 
\end{align}
and the same operators  given in \EQ{eq:AB2}. For these, $C=0$, $L_{\psi_1}=L_{\psi_2}=1,$ while $L_{\varrho}=0.$ Hence, $L_{\varrho}<p_1L_{\psi_1}+p_2L_{\psi_2}$ holds.

Since $L_\varrho$ is neither convex nor concave in $\varrho,$ we will now consider a decomposition of the density matrix to mixed states $\varrho_k$ as given in \EQ{eq:rhodecomp}, instead of a decomposition to pure states.  For such a decomposition, for all $\varrho_k$ the Robertson-Schr\"odinger inequality given in \EQ {eq:RS1} holds. From this fact and with the simple inequality presented in \APP{App:A}, we arrive at
\begin{eqnarray} 
&&\left[\sum_k p_k \va{A}_{\varrho_k}\right]\left [\sum_k p_k \va{B}_{\varrho_k}\right]\nonumber\\
&&\;\;\;\;\;\;\;\;\;\;\;\;\;\;\;\;\;\;\;\;\;\;\;\;\;\;\;\;\;\;\;\;\;\;\;\;\ge\tfrac{1}{4}\left[\sum_k p_k L_{\varrho_k}\right]^2.
\label{eq:Heisenberg2}
\end{eqnarray}
At this point it is important to know that the inequality in \EQ{eq:Heisenberg2} is valid  for any decomposition of the density matrix of the type given in \EQ{eq:rhodecomp}. Moreover, we should remember that the three sums are over the {\it same} decomposition of the density matrix.

Let us try to obtain inequalities with the variance and the QFI. For that, we can choose the decomposition such that 
\begin{equation}
\sum_k p_k \va{B}_{\varrho_k}\label{eq:pkvarB}
\end{equation}
is minimal and equals $F_Q[\varrho,B]/4$ given in \EQ{e2b}. Due to the concavity of the variance we also know that  
\begin{equation}
\sum_k p_k \va{A}_{\varrho_k}\le \va{A}.
\end{equation}
Hence, it follows that for the product of the variance of $A$ and the QFI $F_Q[\varrho,B]$ that
\begin{eqnarray} 
\va{A}_{\varrho}F_Q[\varrho,B]\ge  \left(\sum_k p_k  L_{\varrho_k} \right)^2.\label{eq:RS_improved_with_FQ222}
\end{eqnarray}
In order to use \EQ{eq:RS_improved_with_FQ222}, we need to know the decomposition that minimizes \EQ{eq:pkvarB}. We can have a inequality where 
we do not need to know that decomposition
\begin{eqnarray} 
\va{A}_{\varrho}F_Q[\varrho,B]\ge \left(\min_{\{p_k,\varrho_k\}} \sum_k p_k  L_{\varrho_k} \right)^2.\label{eq:RS_improved_with_FQB}
\end{eqnarray}
On the right-hand side of \EQ{eq:RS_improved_with_FQB}, the bound is defined based on a convex roof.  The right-hand side of \EQ{eq:RS_improved_with_FQB} is not larger than the right-hand side of \EQ{eq:RS_improved_with_FQ222}. We can also see that on the right-hand side of \EQ{eq:RS_improved_with_FQB} there is a minimization over mixed-state decompositions. Based on \SEC{sec:conn-betw-recent}, there is always an optimal pure-state decomposition such that \EQ{eq:pkvarB} is minimal and equals $F_Q[\varrho,B]/4.$ Thus, we can also have a valid inequality with an optimization over pure-state decompositions of the type given in  \EQ{decomp}
\begin{eqnarray} 
\va{A}_{\varrho}F_Q[\varrho,B]\ge \left(\min_{\{p_k,\ket{\psi_k}\}} \sum_k p_k  L_{\psi_k} \right)^2.\label{eq:RS_improved_with_FQ}
\end{eqnarray}
The right-hand side of \EQ{eq:RS_improved_with_FQ} is not smaller than the right-hand side of \EQ{eq:RS_improved_with_FQB}. Hence, in the remaining part of the section we will work with pure-state decompositions rather than mixed-state decompositions.

Let us now try to find a lower bound for the inequality that is easier to compute, while possibly being smaller than the bound in \EQ{eq:RS_improved_with_FQ}. One could first think of using  $\vert L_\varrho \vert^2$ as a lower bound; however, it is not convex in $\varrho,$ as we have discussed. Based on \EQ{eq:defL}, the relation
\begin{equation}
L_{\psi_k} \ge \vert \ex{C}_{\psi_k} \vert \label{eq:ineqLC}
\end{equation}
holds. Based on \EQ{eq:RS_improved_with_FQ} and \EQ{eq:ineqLC}, we can obtain the inequality
\begin{eqnarray} 
&&\va{A}_{\varrho}F_Q[\varrho,B]\ge \left(\min_{\{p_k,\ket{\psi_k}\}} \sum_k p_k\vert \exs{C}_{\psi_k}\vert\right)^2,
\label{eq:RS_improved_with_FQ2}
\end{eqnarray}
Using well-known properties of the absolute value we get 
\begin{equation}
\sum_k p_k\vert \exs{C}_{\psi_k}\vert \ge \left\vert \sum_k p_k \exs{C}_{\psi_k} \right\vert \equiv \vert  \exs{C}_\varrho\vert,
\end{equation}
and with that we arrive at the improved Heisenberg-Robertson uncertainty proved by Fr\"owis \textit{et al.}~\cite{Frowis2015Tighter}
\begin{equation}
\va{A}_{\varrho}F_Q[\varrho,B] \ge \vert  \exs{C}_\varrho\vert^2.
\label{eq:varFQproductUncRel}
\end{equation}
Due to the relation between the variance and the QFI given in \EQ{eq:FQvar}, the left-hand side of \EQ{eq:varFQproductUncRel} is never larger than the left-hand side of the Heisenberg-Robertson uncertainty. 

Based on these, we find the following.

\begin{observation}
\label{obs:conditions}
The improved Heisenberg-Robertson inequality \eqref{eq:varFQproductUncRel} can be saturated only if all of the following conditions are fulfilled.

(i) There is a decomposition $\{p_k,\ket{\psi_k}\}$ that minimizes the weighted sum of the subensemble variances for the operator $B,$ hence
\begin{equation}\label{eq:Fisherroof2}
\frac1 4 F_Q[\varrho,B]= \sum_k p_k \va B_{\psi_k}. 
\end{equation} 
We also need that it maximizes the weighted sum of the subensemble variances for the operator $A,$ and hence
\begin{equation}
\va A_{\varrho}= \sum_k p_k \va A_{\psi_k}. 
\end{equation} 

(ii) If the decomposition maximizes the weighted sum of the subensemble variances for the operator $A$ then \EQ{eq:exApsi} holds. [See explanation after  Eq.~\eqref{eq:Fisherroof}.]
 
(iii) Moreover, \EQ{eq:ineqLC} must be saturated for every $k.$ Hence, the equality 
\begin{equation}
\tfrac{1}{2}\exs{\{A,B\}}_{\psi_k}-\exs{A}_{\psi_k}\exs{B}_{\psi_k}=0
\end{equation}
must hold. In this case, we 
also have
\begin{equation}
\va{(A+B)}_{\psi_k}=\va{A}_{\psi_k}+\va{B}_{\psi_k}.
\end{equation}

(iv) \EQL{eq:Heisenberg2}  is saturated for pure-state decompositions only if for the subensemble variances the equations
\begin{eqnarray}
\va{A}_{\psi_k}&=&\va{A}_{\psi_l},\nonumber\\
\va{B}_{\psi_k}&=&\va{B}_{\psi_l}
\end{eqnarray}
hold for all $k,l.$ (See \APP{App:A}.)

(v) For such an optimal decomposition, for every $k,$ 
\begin{equation}
|\exs{C}_{\psi_k}|=|\exs{C}_{\varrho}|, 
\end{equation}
which is trivially fulfilled if $C$ is a constant. (See \APP{App:A}.) This is the case, for example, if $A$ and $B$ are the position and momentum operators, $x$ and $p,$ of a bosonic mode.
\end{observation}
 
\subsection{Implications for the Cram\'er-Rao bound}

In this section, we will show that the precision of parameter estimation is bounded from below by an expression with the convex roof of the variance.

Let us first define a relevant notion. The error propagation formula is given in \EQ{eq:errprop}. Using the fact that the dynamics is unitary, we have 
\begin{equation}
\vert \partial_{\theta} \exs{A}\vert=\vert \exs{C}\vert,
\end{equation}
where $C$ is defined in \EQ{eq:CAB} (see, e.g., \REF{Frowis2015Tighter}). Hence
\begin{equation} 
\label{eq:errprop2}
\va{\theta}_A=\frac{\va{A}}{\vert \exs{C}\vert^2}.
\end{equation}
Then, the precision of the estimation is bounded as
\begin{equation}
\va{\theta} \ge  \frac{1}{m} \min_{A} \va{\theta}_A,\label{eq:varFQ22}
\end{equation}
where $m$ is the number of independent repetitions. In the large $m$ limit, if certain further conditions are fulfilled, \EQ{eq:varFQ22} can be saturated \cite{Pezze2018Quantum}.

Based on these and on \SEC{subsec:Simple_proof}, we arrive at 
\begin{equation}
\va{\theta}_A\ge \frac{1}{4 \min_{\{p_{k},\ket{\psi_{k}}\}}\bigg[
 \sum_{k}p_{k}
\va{B}_{\psi_k}\bigg]}.
\end{equation}
Using \EQS{eq:errprop2} and \eqref{eq:varFQ22}, we get a lower bound on the precision of parameter estimation
\begin{equation} 
\label{eq:cr}
\va{\theta} \ge \frac1 m \times \frac{1}{4 \min_{\{p_{k},\ket{\psi_{k}}\}}\bigg[
 \sum_{k}p_{k}
\va{B}_{\psi_k}\bigg]}.
\end{equation}
We have just derived a form of the Cram\'er-Rao bound that contains the convex roof of the variance. On the right-hand side of \EQ{eq:cr} we write intentionally the expression with the convex roof, rather than the QFI, to stress that our derivation did not use the formula given in \EQ{eq:qF} for the QFI. Hence, we can see that the Cram\'er-Rao bound in \EQ{eq:cr} can be saturated for von Neumann measurements only if the conditions of \OBS{obs:conditions} are fulfilled for some $A$. 

Note that we did not prove that there is an $A$ for every $B$ and $\varrho$ such that the bound in \EQ{eq:cr} can be saturated, which would be necessary to prove that the Cram\'er-Rao bound can be reached.

Note also that we did not consider POVM measurements, which would be the more general case \cite{Braunstein1994Statistical}. However, it is known that it is always possible to saturate  the Cram\'er-Rao bound by von Neumann measurements \cite{Paris2009QUANTUM}. 

\subsection{Sufficient condition for saturating the bound}

In this section, for completeness, we present a concise sufficient condition that \EQ{eq:varFQproductUncRel} is saturated. Similar statements have been discussed in \REFS{Hotta2004Quantum,Escher2012Quantum_arxiv,Frowis2015Tighter,Toth2020Activating}. This is relevant for us, since it is connected to the conditions for saturation given in \OBS{obs:conditions}. We use the theory of the symmetric logarithmic derivative described in \SEC{sec:Important_properties}.

\begin{observation} 
\label{obs:saturation}
If the equality
\begin{equation}
i[\varrho,B] = \tfrac{1}{2} \left\{ \varrho, c A\right\},\label{eq:comm_anticomm}
\end{equation}
holds, then \EQ{eq:varFQproductUncRel} is saturated. Here, $c\ne 0$ is a real constant.
\end{observation}

{\it Proof.} \EQL{eq:comm_anticomm} implies that $cA$ equals the symmetric logarithmic derivative $\mathcal{L};$ see \EQ{eq:comm_anticomm2}. Let us then substitute $cA$ by $\mathcal{L}$ in \EQ{eq:varFQproductUncRel}. 
Then it follows that \cite{Braunstein1994Statistical,Braunstein1996Generalized}
\begin{equation}
F_Q[\varrho,B]= \langle \mathcal{L}^2 \rangle_{\varrho} = c^2 (\Delta A)^2_{\varrho},
\end{equation}
and moreover simple algebra yields
\begin{equation}
\langle  i[A,B] \rangle_{\varrho} = \frac1 c \trace(i[\mathcal{L},B]\varrho)= \frac1 c \langle \mathcal{L}^2 \rangle_{\varrho}= c(\Delta A)^2_{\varrho}.\label{eq:Lcomm}
\end{equation}
In the last equality in \EQ{eq:Lcomm} we used \EQ{eq:L0}. Consequently, the left-hand side and the right-hand side of \EQ{eq:varFQproductUncRel} are equal, and the state saturates the inequality. Moreover, the two terms of the product on the left-hand side of \EQ{eq:varFQproductUncRel} are equal to each other if $c=1,$ that is, $\va{A}=F_Q[\varrho,B].$  $\qed$

\OBS{obs:saturation} is related to the known relation
\begin{equation}
\va{\mathcal L}F_Q[\varrho,B]=\vert\langle  i[\mathcal L,B] \rangle_{\varrho} \vert^2,\label{eq:factors}
\end{equation}
where $\mathcal L$ is defined in \EQ{eq:L}, and on the left-hand side of \EQ{eq:factors}, the two terms in the product are equal to each other.

Note that as a consequence, the equality in \EQ{eq:comm_anticomm} implies  that conditions in \OBS{obs:conditions} are fulfilled. Moreover, based on \OBS{obs:saturation}, we can find additional constraints on the subsensemble variances given in \OBS{obs:conditions}.  If we compute the trace of both sides of \EQ{eq:comm_anticomm} we arrive at
\begin{equation}
\exs{A}_\varrho=0.
\end{equation}
From \OBS{obs:conditions} (ii) follows that there is a similar statement for all subensembles
\begin{equation}
\exs{A}_{\psi_k}=0.
\end{equation}
Moreover, based on \OBS{obs:conditions} (iii), we obtain a relation about the commutator of $A$ and $B$ as
\begin{equation}
\exs{\{A,B\}}_{\psi_k}=0,
\end{equation}
which is again valid for all subensembles.

\section{Improvement on the Robertson-Schr\"odinger inequality based on a concave roof over decompositions}
\label{sec:RS}

In this section, we show an improvement of the Robertson-Schr\"odinger inequality. We start from the fact that \EQ{eq:Heisenberg2} is valid for any decomposition of the density matrix to mixed components $\varrho_k.$ Hence, due to the concavity of the variance follows that
\begin{equation}
\va{A} \va{B} \ge\frac{1}{4}\left(\sum_k p_k L_{\varrho_k}\right)^2.
\label{eq:Heisenberg2BB}
\end{equation}
Based on these, we can find the following.

\begin{observation} For quantum states, the following inequality holds
\label{obs:vavaL_ineq}
\begin{eqnarray} 
\va{A}_{\varrho}\va{B}_{\varrho}\ge \frac 1 4 \left( \max_{\{p_k,\varrho_k\}} \sum_k p_k  L_{\varrho_k} \right)^2,\label{eq:RS2B}
\end{eqnarray}
where $L_{\varrho}$ is defined in \EQ{eq:defL}. On the right-hand side of \EQ{eq:RS2B}, we have a concave roof. The relation in \EQ{eq:RS2B} is saturated by all single-qubit mixed states, and it is stronger than the Robertson-Schr\"odinger inequality given in \EQ{eq:RS1}.
\end{observation}

{\it Proof.} By taking the maximum of the bound on the right-hand side of \EQ{eq:Heisenberg2BB} over mixed-state decompositions, we arrive at the inequality \EQ{eq:RS2B}. 

Let us examine the single-qubit case in detail. Let us take the operators
\begin{eqnarray}
A &=& \sigma_x,\nonumber\\
B &=& \cos \alpha \sigma_x + \sin \alpha \sigma_y,\label{eq:AB}
\end{eqnarray}
which is the most general case, apart for trivial rotations of the coordinate system. We characterize the state by the Bloch vector elements $\exs{\sigma_l}$ for $l=x,y,z.$ Substituting these in the bound in the Robertson-Schr\"odinger inequality given in  \EQ{eq:RS1} we obtain for pure states
\begin{eqnarray}
\frac1 4 \vert L_{\psi}\vert^2&=&
 \left[ \cos \alpha (\Delta \sigma_x)^2_{\psi} - \sin\alpha \langle \sigma_x \rangle_{\psi}\langle \sigma_y \rangle_{\psi}\right]^2 \nonumber\\ &+& \sin^2\alpha \langle \sigma_z \rangle_{\psi}^2.
\label{eq:Lpure}
\end{eqnarray}
Substituting $\langle \sigma_z \rangle_{\psi}^2 = 1- \langle \sigma_x \rangle_{\psi}^2 - \langle \sigma_y \rangle_{\psi}^2$
into \EQ{eq:Lpure} we arrive at
\begin{eqnarray}
\label{eq:15}
\frac1 4 \vert L_{\psi}\vert^2&=&(\Delta \sigma_x)^2_{\psi} \bigg[ \cos^2 \alpha (\Delta \sigma_x)^2 \nonumber\\ &-& \sin 2\alpha \langle \sigma_x \rangle_{\psi}\langle \sigma_y \rangle_{\psi} + \sin ^2 \alpha (\Delta \sigma_y)^2\bigg].
\end{eqnarray}
Simple algebra leads to 
\begin{eqnarray}
\frac1 4 \vert L_{\psi}\vert^2&=&(\Delta A)^2_{\psi} (\Delta B)^2_{\psi}.
\end{eqnarray}
Hence, all pure states saturate \EQ{eq:RS2B}.

We will now show that for every single-qubit mixed state and every single-qubit operator the inequality in \EQ{eq:RS2B} is saturated. If we can find a decomposition $\left\{ p_k,  \varrho_k \right\}$, such that 
\begin{equation}
( \Delta X )_{\varrho}^2 = ( \Delta X )_{\varrho_k}^2
\end{equation}
for $X = A,B$ and all $k$, then this is sufficient to have equality in \EQ{eq:RS2B}. The following decomposition has this property. We imagine the Bloch sphere with a vector representing an arbitrary $\varrho$ and a straight line that goes through $\varrho$ and that is parallel to the $z$ axis. All states along this line within the Bloch sphere have the same expectation values for $\sigma_x$ and $\sigma_y$ as $\varrho,$ hence also the same variances. Therefore, we choose a decomposition of $\varrho$ with the pure states at the points where the line intersects with the Bloch sphere. With this, we have equality in  Eq.~(\ref{eq:RS2B}).

Let us now prove that the Robertson-Schr\"odinger inequality in Eq.~(\ref{eq:RS2B}) is stronger than  the Robertson-Schr\"odinger inequality given in \EQ{eq:RS1}. First, the right-hand side of \EQ{eq:RS2B} is never smaller than the right-hand side of the Robertson-Schr\"odinger inequality given in \EQ{eq:RS1}. This is evident since one of the possible decompositions is 
\begin{equation}
p_1=1,\quad\quad\quad\varrho_1=\varrho.
\end{equation}
Second, let us now consider a concrete example when the bound in \EQ{eq:RS2B} is higher than the bound in the Robertson-Schr\"odinger inequality given in \EQ{eq:RS1}. For instance, let us consider the complitely mixed state 
\begin{equation}
\varrho=\openone/2 \label{eq:denmat}
\end{equation}
and the operators given in \EQ{eq:AB}. Since we found that the inequality in Eq.~(\ref{eq:RS2B}) is saturated by all single-qubit mixed states, 
it is also saturated for the state given in \EQ{eq:denmat} and the right-hand side of Eq.~(\ref{eq:RS2B}) equals $1.$ The right-hand side of 
the Robertson-Schr\"odinger inequality given in Eq.~(\ref{eq:RS1}) is zero. $\qed$

Let us consider now higher dimensional systems. Here, not all pure states saturate the inequality given in \EQ{eq:RS1}. As an example, let us consider qutrit states, and the operators
\begin{eqnarray}
A&=&J_x,\nonumber\\
B&=&J_y.
\end{eqnarray}
First we show a simple method that never gives a bound lower than the bound in \EQ{eq:RS1}, and often it gives a higher bound.  Let us consider the bound based on \EQ{eq:Heisenberg2BB} using the eigendecomposition of $\varrho$ given in \EQ{eq:rho_eigdecomp}:
\begin{eqnarray} 
\va{A}_{\varrho}\va{B}_{\varrho}\ge \frac 1 4 \left(\sum_k \lambda_k  L_{\ket k}\right)^2 .\label{eq:eig1}
\end{eqnarray} 
Since $L_\varrho$ is not convex in $\varrho,$ the bound in the inequality given in \EQ{eq:eig1} might be smaller than the bound in the Robertson-Schr\"odinger relation given in \EQ{eq:RS1}. 

We will now present a relation for which the bound is never smaller than in \EQ{eq:RS1}. Let us consider the unnormalized states
\begin{equation}
\sigma_{k}=\varrho-\lambda_k\ketbra{k}
\end{equation}
for $k=1,2,3.$  Here, $\sigma_{k}$ is a mixture of the two basis vectors orthogonal to $\ket{k}.$ Then we define the probabilities and normalized states
\begin{equation}
p_k={\rm Tr}(\sigma_k),\quad\quad\quad\varrho_k=\sigma_k/p_k.
\end{equation}
Using $\varrho_k,$ we decompose the density matrix as 
\begin{equation}
\varrho=\lambda_k \ketbra{k}+p_k \varrho_k,
\end{equation}
where $k\in\{1,2,3\}.$
With these we define the quantity
\begin{equation}
\tilde L_k=\lambda_k L_{\ket{k}}+p_k  L_{\varrho_{k}}.
\end{equation}
Then, we obtain an inequality
\begin{eqnarray} 
\va{A}_{\varrho}\va{B}_{\varrho}\ge \frac 1 4 K^2.\label{eq:eig12}
\end{eqnarray} 
where the variable in the bound is defined as 
\begin{equation}
K=\max\left(\sum_k \lambda_k  L_{\ket k},  \tilde L_1, \tilde L_2, \tilde L_3, L_{\varrho}\right),\label{eq:K}
\end{equation}
and $\max(a_1,a_2,a_3,....)$ denotes the maximum of $a_k.$ Since $K\ge L_{\varrho},$ the bound in \EQ{eq:eig12} is never smaller than the bound in the Robertson-Schr\"odinger inequality given in \eqref{eq:RS1}. We will see that it is often larger. We considered all the possible ways to group  the eigenvectors into groups and form mixed states from them. Such ideas can straightforwardly be generalized to larger dimensions, where we need to consider more partitions of the eigenvectors. 

\begin{figure}[t!]
\centerline{ \epsfxsize1.65in \epsffile{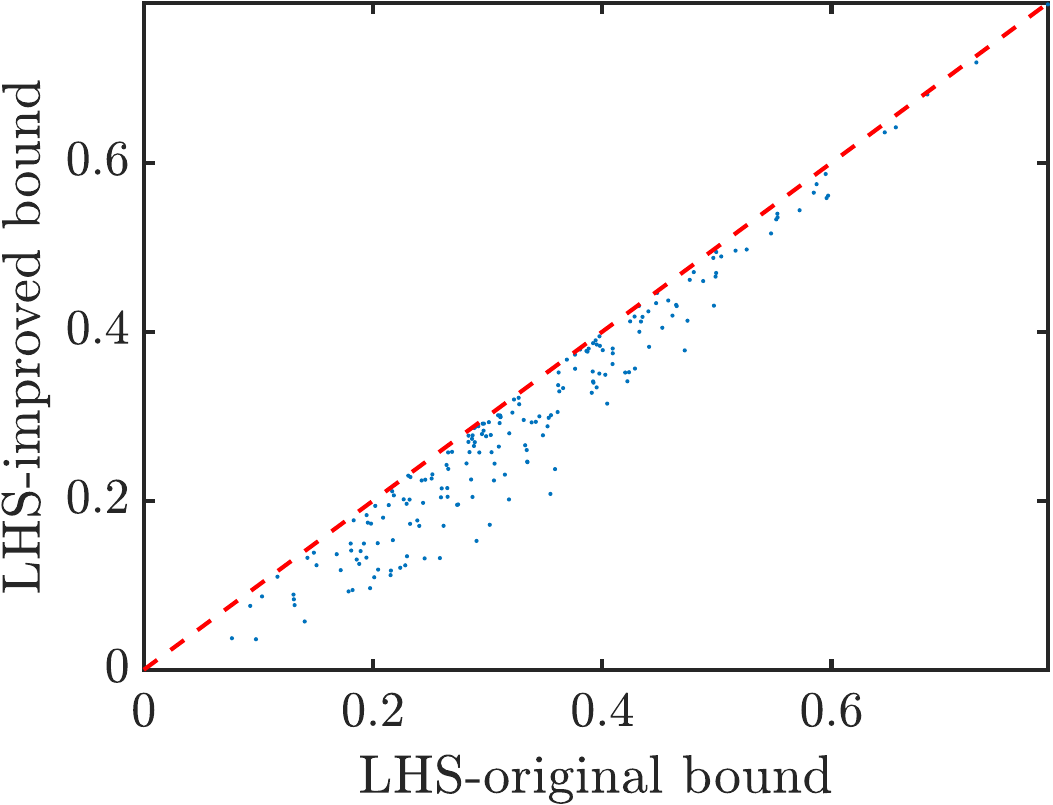}\hskip0.3cm
\epsfxsize1.65in \epsffile{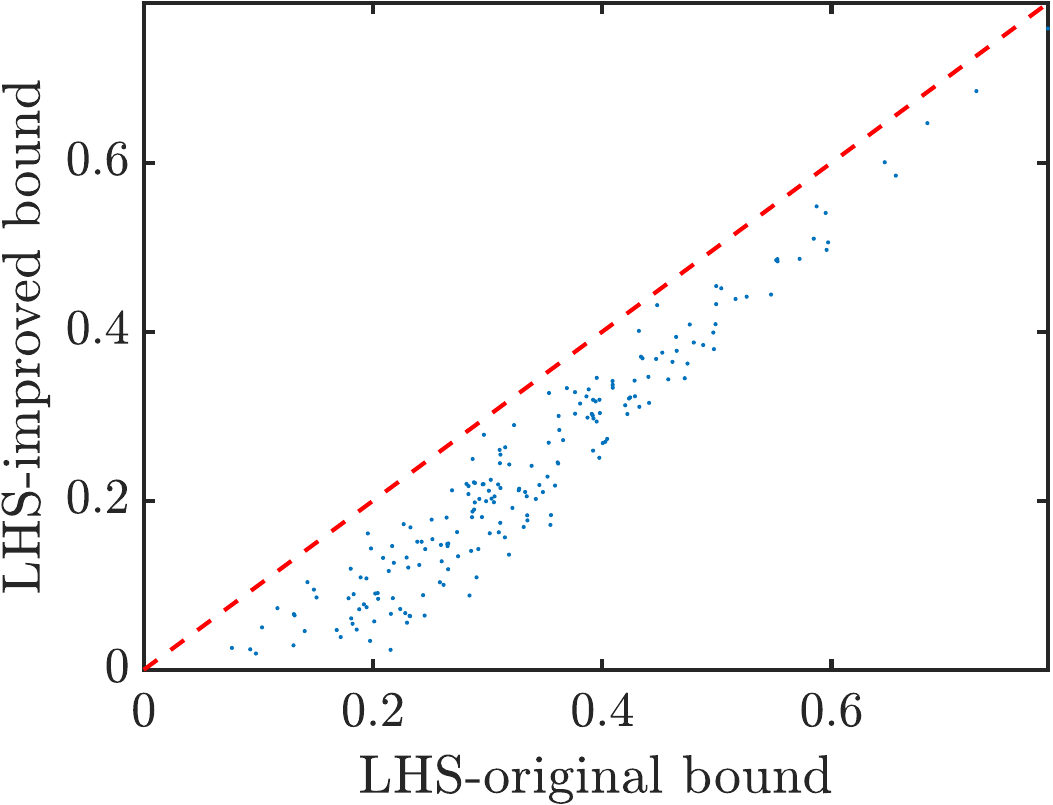}}
\hskip0.5cm (a) \hskip4cm (b)
\caption{(a) The left-hand side (LHS) minus the right-hand side (RHS) for the  Robertson-Schr\"odinger inequality \eqref{eq:RS1} vs. LHS-RHS for \EQ{eq:RS2B}, taking the eigendecomposition of the density matrix and using the inequality given in \EQ{eq:eig12}. Points that are below the dashed line correspond to quantum states for which the bound improved. Even if our method is simple, the improvement is significant. We generated $200$ random states. (b) The same for the concave roof obtained numerically.}
\label{fig:rs}
\end{figure}

Next, we will test the uncertainty relation given in \EQ{eq:RS2B} numerically. We generate random single-qutrit states \cite{Zyczkowski2001Induced}. We calculate the usual bound in the Robertson-Schr\"odinger inequality given in \EQ{eq:RS1}. We test the simple method given in \EQ{eq:eig12} that give improved bounds.  The results can be seen in \FIG{fig:rs}(a). Note that typically we do not find the best possible bound but still this simple technique often leads to an improvement and the new bound is significantly larger than the old one. There are also numerical methods to compute the concave roof in \EQ{eq:RS2B}, described in \APP{sec:numerical}. Based on that we carry out numerical optimization over mixed-state decompositions. The results can be seen in \FIG{fig:rs}(b). Even if the numerical search might not find the global  maximum, but something smaller, 
we found a valid lower bound on the left-hand side of \EQ{eq:RS2B}.

\section{Uncertainty relations with several variances and the QFI}
\label{sec:several_var}

In this section, we derive uncertainty relations with the QFI, and one or more variances. This provides a lower bound on the QFI based on variances of angular momentum operators.

\subsection{Sum of two variances} 
\label{sec:Sum of two variances}
Ideas similar to the ones in \SEC{sec:convroof} work even if we have uncertainty relations that are the sum of two variances. For example, for a continuous variable system 
\begin{equation}
\va{x}+\va{p}\ge 1
\end{equation}
 holds, where $x$ and $p$ are the position and momentum operators. This must be valid for any state, including pure states. Hence, for any decompositions of the density matrix it follows that
\begin{equation}
\sum_k p_k \va{x}_{\psi_k}+\sum_k p_k \va{p}_{\psi_k}\ge 1
\end{equation}
For one of the two operators, say, for $p,$ we can choose the decomposition that leads to the minimal value for the average variance, i.e., the QFI over four. Then, since $\sum_k p_k \va{x}_{\psi_k}\le \va{x}$ holds, it follows that
\begin{equation}
\va{x}+\tfrac{1}{4}F_Q[\varrho,p]\ge 1.
\end{equation}

Note that this could be obtained more directly from the uncertainty relation in \EQ{eq:varFQproductUncRel} using 
\begin{equation}
X+Y \ge 2 \sqrt{XY}, \label{eq:XYsqrt}
\end{equation}
for $X,Y\ge 0,$ but we intended to demonstrate 
the key idea of the next sections.

\subsection{Lower bound on the QFI} 
\label{sec:3varfisher}

Similar reasoning works for the uncertainty relations for the sum of three variances. Let us start from the relation for pure states
\begin{equation}\label{eq:JxyzN2b}
\va{J_x}+\va{J_y}+\va{J_z} \ge j,
\end{equation}
where $J_l$ are the spin components fulfilling
\be
J_x^2+J_y^2+J_z^2=j(j+1)\openone.
\end{equation}
Due to the concavity of the variance, the inequality in \EQ{eq:JxyzN2b} holds also for mixed states. We can improve this relation. From the inequality given in \EQ{eq:JxyzN2b}, following the ideas of \SEC{sec:Sum of two variances}, we arrive at 
\begin{equation}
\label{eq:Jxyzvar2}
\va{J_x}+\va{J_y}+\tfrac{1}{4}F_Q[\varrho,J_z]\ge j.
\end{equation}

\EQL{eq:Jxyzvar2} is a stronger relation than \EQ{eq:JxyzN2b}. The left-hand side of \EQ{eq:Jxyzvar2} is never smaller than the left-hand side of \EQS{eq:JxyzN2b}. The difference between the two is given in \EQ{eq:vaminusFQ}. This quantity has been studied in \REF{Toth2017Lower}. It is zero for pure states and largest for the state 
\begin{equation}
\frac1{ 2}\left(\ketbra{-j}_z+\ketbra{+j}_z\right).\label{eq:pmz}
\end{equation}
For this case, for the $z$-component of the spin we have $\va{J_z}=j^2$ and $F_Q[\varrho,J_z]=0,$ while for the variance of the $x$-component we have $\va{J_x}=j/2.$ 
Thus, the state given in \EQ{eq:pmz} saturates the inequality in \EQ{eq:Jxyzvar2}.

Based on these, we arrive at the following observation.

\begin{observation} 
For a spin-$j$ particle, the following inequality bounds from below the metrological usefulness of the state 
\be\label{eq:FQbound}
F_Q[\varrho,J_z]\ge 4j-4\va{J_x}-4\va{J_y}=:B_{FQ}.
\end{equation}
\end{observation}

Let us now examine whether the bound in \EQ{eq:FQbound} can be improved. It is known that \EQ{eq:JxyzN2b}  is saturated by all pure 
SU(2) coherent states or spin-coherent states, which are defined as
\begin{equation}
\ket{s}=U\ket{+j}_z, \label{eq:spincoherent}
\end{equation}
where the unitary is given as
\begin{equation}
U=e^{-i\vec{c}\vec{J}},
\end{equation}
where $\vec{c}$ is a three-vector of numbers and $\vec{J}=(J_x,J_y,J_z).$ Hence,  the inequality given in \EQ{eq:FQbound} is also saturated by all such states, and the bound is optimal. 

The inequality in \EQ{eq:FQbound} bounds the QFI from below based on variances. Such a bound can be very useful in some situations, since we do not need to carry out a metrological task to get information about $F_Q[\varrho,J_z].$ Let us now consider some quantum states, and compare the bound given by \EQ{eq:FQbound} to the QFI of those states. Our first example will be planar squeezed states \cite{He2011Planar}. Such states saturate the uncertainty relation
\be\label{eq:planar}
\va{J_x}+\va{J_y}\ge C_j,
\end{equation}
where for the bound
\begin{equation}
C_{\frac{1}{2}}=\tfrac{1}{4},\quad C_{1}=\tfrac{7}{16}
\end{equation}
holds, while for higher $j$'s the bound is obtained numerically  in \REF{He2011Planar}. Note that planar squeezed states minimize  the left-hand side of \EQ{eq:planar} such that their mean spin is not zero.  Thus, they are different from the states that minimize $\exs{J_x^2}+\exs{J_y^2}.$

\begin{figure}[t!]
\centerline{ 
\epsfxsize2.5in \epsffile{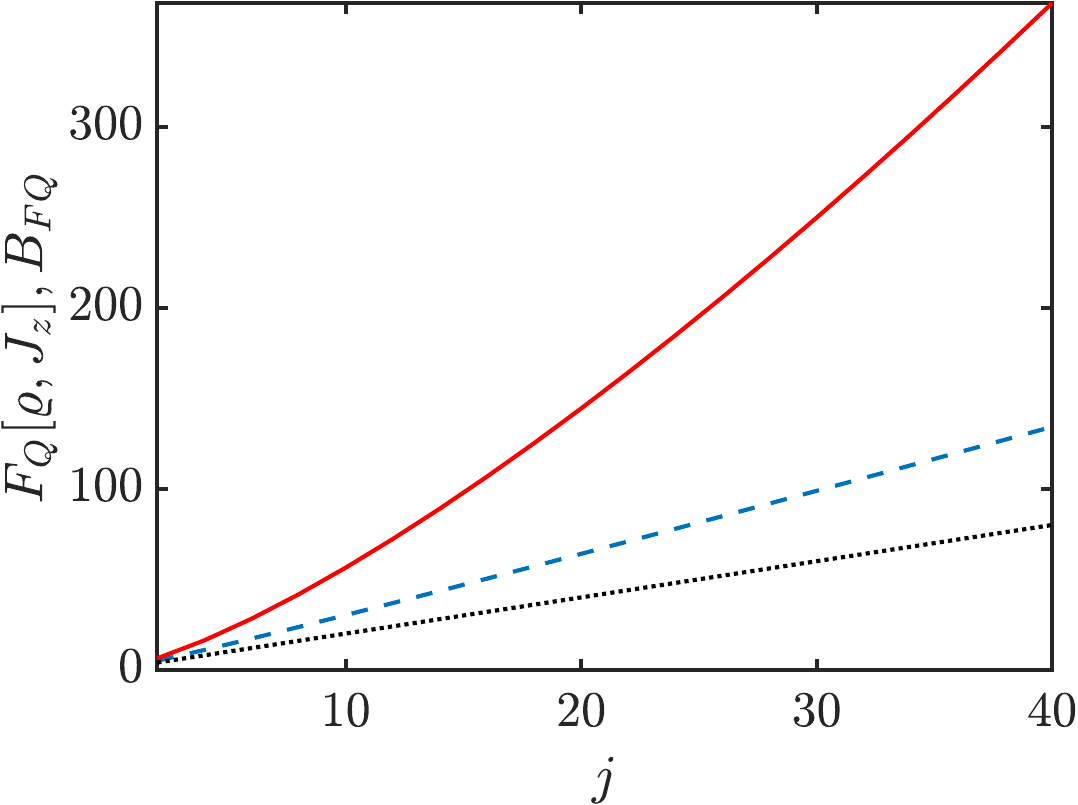}
}
\caption{Planar squeezed states. (solid) The QFI and (dashed) our lower bound $B_{FQ}$ given in \EQ{eq:FQbound} for planar squeezed states for a range of $j.$ (dotted) As a reference, we plot the QFI corresponding to the state fully polarized in the $x$-direction.}
\label{fig:planar}
\end{figure}

\begin{figure}[t!]
\centerline{ 
\epsfxsize2.5in \epsffile{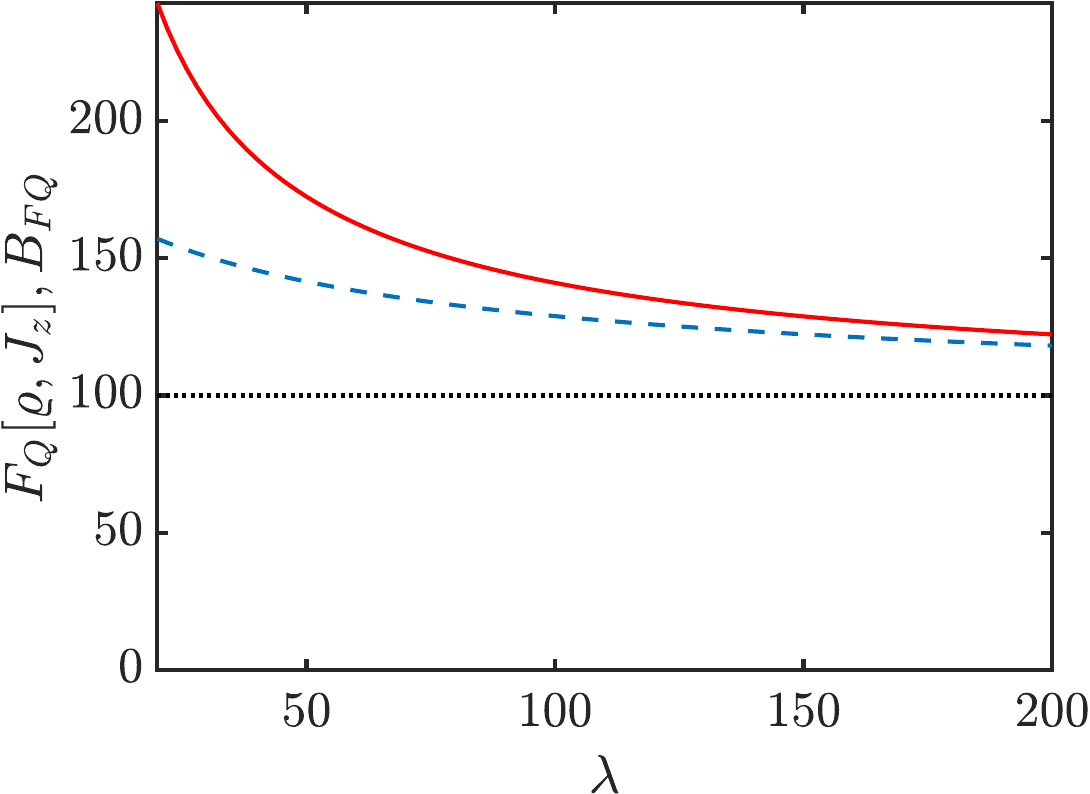}}
\caption{ Spin-squeezed states. (solid) The QFI and (dashed) our lower bound $B_{FQ}$  defined in  \EQ{eq:FQbound} for spin-squeezed states $j=50.$ The spin-squeezed states are obtained as the ground states of \EQ{eq:Hsq} for a range of $\lambda.$   (dotted) As a reference, we plot the QFI corresponding to the state fully polarized in the $x$ direction.}
\label{fig:spinsq}
\end{figure}

Based on the inequalities given in \EQS{eq:FQbound} and \eqref{eq:planar}, for planar squeezed states we have 
\begin{equation}\label{eq:FQbound2}
F_Q[\varrho,J_z]\ge B_{FQ} =  4(j-C_j).
\end{equation}
In \EQ{eq:FQbound2}, the value of $B_{FQ}$  approaches $4j$  since for large $j$ we have \cite{He2011Planar}
\begin{equation}
C_j\ll j. 
\end{equation}
In \FIG{fig:planar}, we plotted the QFI vs. our lower bound for planar squeezed states for various $j$'s. 

We will present another class of states for which our bound on the QFI can be useful. We will consider the state $\ket{j}_x$ squeezed in the $y$-direction. Spin-squeezed states can be obtained as the ground states of the Hamiltonian \cite{Sorensen2001Entanglement}
\begin{equation}\label{eq:Hsq}
H_{{\rm sq}}(\lambda)=J_y^2-\lambda J_x.
\end{equation}
For $\lambda=\infty,$ the ground state is $\ket{j}_x$, the state fully polarized in the $x$-direction. For $0<\lambda<\infty,$ it is a state spin squeezed along the $y$-direction. When the state becomes squeezed along the $y$-direction, the sum of the two variances in \EQ{eq:FQbound} starts to decrease. Then, due to \EQ{eq:FQbound}, the QFI has to increase and the state becomes more useful for metrology. In \FIG{fig:spinsq}, we plotted the right-hand side and the left-hand side of \EQ{eq:FQbound} for a range of $\lambda$ for $j=50.$ Our lower bound is quite close to the QFI for states with an almost maximal spin.

Next, we will determine what the largest precision is for SU(2) coherent states or spin-coherent states defined in \EQ{eq:spincoherent}. It is easy to show that $F_Q[\varrho,J_z]$ is maximal for $\ket{j}_x,$ the SU(2) coherent state pointing into in the $x$-direction. For that state, we have
\begin{equation}
F_Q[\ket{j}_x,J_z]=2j,\quad\va{J_x}=0,\quad \va{J_y}=j/2.
\end{equation}
Due to the convexity of the QFI, for the mixtures of SU(2) coherent states
\begin{equation}
\label{eq:SU2m}
F_Q[\varrho_{{\rm SU}(2){-\rm mixture}},J_z]\le 2j
\end{equation}
holds. Any state that violates \EQ{eq:SU2m} is more useful metrologically than a mixture  of SU(2) coherent states. Both in \FIG{fig:planar} and in \FIG{fig:spinsq}, we plot a line corresponding to the bound in the inequality given in \EQ{eq:SU2m}.

Finally, let us generalize these ideas to more than three operators. Let us consider the following relation for pure states
\begin{equation}\label{eq:Gkequality}
\sum_{n=1}^{d^2-1} \va{G_n} = 4j,
\end{equation}
where $G_n$ are the SU($d$) generators fulfilling
\begin{equation}
{\rm Tr}(G_kG_l)=2\delta_{kl},
\end{equation}
and, $d=2j+1$ is the dimension of the qudit  (see e.g., \REF{Vitagliano2011Spin}). Due to the concavity of the variance it follows that for mixed states \cite{Vitagliano2011Spin}
\be\label{eq:varGk}
\sum_{n=1}^{d^2-1} \va{G_n} \ge 4j,
\end{equation}
We can even have a better relation based on the discussion before.

\begin{observation}
For a spin-$j$ particle, the following inequality bounds from above
the metrological usefulness of the state 
\begin{equation}\label{eq:varGk_imrpoved}
\tfrac{1}{4} F_Q[\varrho,G_1]+\sum_{n=2}^{d^2-1} \va{G_n} \ge 4j.
\end{equation}
\end{observation}

\section{Alternative derivation based on convexity arguments}
\label{sec:alt}

In this section, we present a simple idea that can be used to rederive some of the previous results. The derivation becomes much shorter, while the conditions for saturating the inequalities are not so easy to obtain. We also derive further inequalities with the variance and the QFI. Part of the section is a summary of already existing results, which we are connecting to the methods of the paper.

\begin{observation}
\label{obs:convroof}
Let us consider a relation
\begin{equation}
\va{A}_\varrho \ge g(\varrho),\label{eq:vag}
\end{equation}
which is true for pure states. If $g(\varrho)$ is convex in density matrices, then
\begin{equation}
\frac 1 4 F_Q[\varrho,A] \ge g(\varrho) \label{eq:vag22}
\end{equation}
holds for mixed states. If  $g(\varrho)$ is not convex in $\varrho,$  the inequality
\begin{equation}
\frac 1 4 F_Q[\varrho,A] \ge \min_{\{p_k,\ket{\psi_k}\}}  \sum_k p_k g(\ket{\psi_k}) \label{eq:vag22bfq}
\end{equation}
\end{observation}
still holds.

{\it Proof.}--On the left-hand side of the inequality in \EQ{eq:vag22} there is the QFI of $\varrho$ over four. Based on \EQ{e2b}, we know that it is a convex roof, that is, the largest convex function that equals $\va{A}_\varrho$ for all pure states. If $g(\varrho)$ is convex in $\varrho,$ then, on the right-hand side of Eq.~(\ref{eq:vag22}), there is an expression that is never larger than the left-hand side for pure states. Even if $g(\varrho)$ is not convex in $\varrho,$ then the right-hand side of Eq.~(\ref{eq:vag22bfq}) is still convex in $\varrho.$ $\qed$

Using \OBS{obs:convroof}, the inequality with the product of the variance and the QFI given in \EQ{eq:RS_improved_with_FQ} can be proved as follows. 
We will use the ideas of \REF{Giovanetti2003Characterizing} to convert relations with the product of uncertainties to relations with the sum of uncertainties.
Based on \EQ{eq:XYsqrt}, we know that
\begin{equation}
\alpha \va{A} + \beta \va{B} \ge 2\sqrt{\alpha\beta} \sqrt{ \va{A} \va{B} }\label{eq:XYsqrt2}
\end{equation}
holds for all $\alpha,\beta\ge0.$ From the product uncertainty relation given in \EQ{eq:RS1} and from \EQ{eq:XYsqrt2} follows that for the weighted sum of the variances
\begin{equation}
\alpha \va{A} + \beta \va{B} \ge \sqrt{\alpha\beta} L_{\varrho}. \label{eq:ABsqrt}
\end{equation}
holds for all $\alpha,\beta\ge0.$ From \EQ{eq:ABsqrt}, we can get a relation with a quantity on the right-hand side that is convex in the density matrix 
\begin{equation}
\alpha \va{A} + \beta \va{B} \ge \sqrt{\alpha\beta} \min_{\{p_k,\ket{\psi_k}\}} \sum_k p_k  L_{\psi_k}. \label{eq:ABsqrt2}
\end{equation}
Let us rewrite \EQ{eq:ABsqrt2} as
\begin{equation}
\beta \va{B} \ge \sqrt{\alpha\beta} \min_{\{p_k,\ket{\psi_k}\}} \left(\sum_k p_k  L_{\psi_k} \right)-\alpha \va{A} . \label{eq:ABsqrt3}
\end{equation}
Now on the left-hand side we have a variance while the right-hand side is convex in the state. Using Observation 5, we arrive at 
\begin{equation}
\beta \frac 1 4 F_Q[\varrho,B] \ge \sqrt{\alpha\beta} \min_{\{p_k,\ket{\psi_k}\}} \left(\sum_k p_k  L_{\psi_k} \right)-\alpha \va{A} . \label{eq:ABsqrt4}
\end{equation}
Hence, we arrive at a relation with the wighted sum of the variance and the QFI
\begin{equation}
\alpha \va{A}+\beta \frac 1 4 F_Q[\varrho,B] \ge \sqrt{\alpha\beta} \min_{\{p_k,\ket{\psi_k}\}} \left(\sum_k p_k  L_{\psi_k} \right). \label{eq:ABsqrt4B}
\end{equation}
From the fact that the inequality in \EQ{eq:ABsqrt4} holds for all $\alpha,\beta\ge0$ follows the inequality with the product of the variance and the QFI given in \EQ{eq:RS_improved_with_FQ}.

Using \OBS{obs:convroof}, the uncertainty relation with two variances and the QFI given in \EQ{eq:FQbound} can be proved as follows. The uncertainty relation with three variances in \EQ{eq:JxyzN2b} can be rewritten as
\begin{equation}
\va{J_x}\ge j-\va{J_y}-\va{J_z},
\end{equation}
which is of the form given in \EQ{eq:vag}, since its right-hand side is convex in $\varrho$ and the left-hand side is a variance.
Hence, the inequality in \EQ{eq:FQbound} can be rederived.

Next, we prove a general bound on the metrological usefulness of a quantum state based on its spin length using \OBS{obs:convroof}.

\begin{observation}
The metrological usefulness of a state is bounded with the spin-length as
\begin{equation}
F_Q[\varrho,J_x]\ge 4jF_j(\ex{J_z}/j).\label{eq:FQFjbound}
\end{equation}
where  $F_j(X)$ is a convex function defined as
\begin{equation}
F_j(X)=\min_{\varrho:\ex{J_z}=Xj} \frac{\va{J_x}}{j}.\label{eq:Fjdef}
\end{equation}
In particular, if $\ex{J_z}\ne 0$ then $F_Q[\varrho,J_x]>0.$ 
\end{observation}

{\it Proof.} For the components of the angular momentum for a particle with spin-$j$ \cite{Sorensen2001Entanglement}
\begin{equation}
\va{J_x}\ge jF_j(\ex{J_z}/j)
\end{equation}
holds.  Using \OBS{obs:convroof}, we can obtain an inequality for the QFI
\begin{equation}
\frac 1 4  F_Q[\varrho,J_x]\ge jF_j(\ex{J_z}/j).
\end{equation}
Based on the definition in \EQ{eq:Fjdef}, it is clear that if $\ex{J_z}>0$ then $F_j(\ex{J_z}/j)>0.$ Hence, based on the relation in \EQ{eq:FQFjbound} follows that $F_Q[\varrho,J_x]>0.$ Thus, if the $z$-component of the angular momentum has a non-zero expectation value then the  state can be used for metrology with the Hamiltonian $J_x.$ $\qed$

Usually, the function $F_j(X)$ is computed by looking for the ground state $\ket{\Psi_{\lambda,\lambda_2}}$ of the Hamiltonian \cite{Sorensen2001Entanglement}
\begin{equation}
H_{\lambda,\lambda_2}=J_x^2-\lambda J_z - \lambda_2 J_x,\label{eq:Hlambdalabda2}
\end{equation}
where $\lambda$ and $\lambda_2$ play the role of Lagrange multipliers. In particular, we need the ground state of the Hamiltonian given in \EQ{eq:Hlambdalabda2} for which $\ex{J_z}$ equals a given value and $\va{J_x}$ is minimal, that is,
\begin{equation}
F_j(X)=\min_{\lambda,\lambda_2:\ex{J_z}_{\ket{\Psi_{\lambda,\lambda_2}}}=Xj} \va{J_x}_{\ket{\Psi_{\lambda,\lambda_2}}}/j.\label{eq:Fjbound}
\end{equation}
In this way the minimization is over two real parameters, rather than over a quantum state.  Such a calculation has been used to obtain a lower bound on the variance $\va{J_x},$ if the expectation value $\ex{J_z}$ is constrained to be a given constant \cite{Sorensen2001Entanglement}.  For an integer $j,$ the state minimizing $\va{J_x}$ has $\ex{J_x}=0,$ thus $\lambda_2$ can be omitted from the Hamiltonian in \EQ{eq:Hlambdalabda2}. Later it has also been shown that with such a procedure we get a lower bound on $\frac 1 4  F_Q[\varrho,J_x]$ \cite{Apellaniz2017Optimal}.

\begin{observation}
\label{obs:convevroof_multiterm}
Let us consider a relation
\begin{equation}
\sum_{n=1}^{N_A} \va{A_n}_\varrho \ge g(\varrho),
\end{equation}
which is true for pure states with some $A_n$ operators. Here $N_A$ is the number of $A_n$ operators we consider. If $g(\varrho)$ is convex in density matrices, then
\begin{equation}
I(\{A_n\}_{n=1}^{N_A},\varrho) \ge g(\varrho)
\end{equation}
holds for mixed states, where we define
\begin{equation}
I(\{A_n\}_{n=1}^{N_A},\varrho)= \min_{\{p_k,\ket{\psi_k}\}}\sum_k p_k \sum_{n=1}^{N_A} \va{A_n}_{\psi_k}.\label{eq:infsum}
\end{equation}
If  $g(\varrho)$ is not convex in $\varrho$ then the inequality with a convex roof 
\begin{equation}
I(\{A_n\}_{n=1}^{N_A},\varrho) \ge \min_{\{p_k,\ket{\psi_k}\}}  \sum_k p_k g(\ket{\psi_k}) \label{eq:vag22bbb}
\end{equation}
still holds.
\end{observation}

{\it Proof.} The proof is analogous to that of \OBS{obs:convroof}. $\qed$

For a single operator 
\begin{equation}
I(\{A\},\varrho)= \frac1 4  F_Q[\varrho,A]
\end{equation}
holds. For two or more operators, it is clear that $I(\{A_n\}_{n=1}^{N_A},\varrho) $ can be larger than the sum of the corresponding QFI terms 
\begin{equation}
I(\{A_n\}_{n=1}^{N_A},\varrho) \ge \frac1 4 \sum_{n=1}^{N_A} F_Q[\varrho,A_n].
\end{equation}

Note that based on the ideas of \SEC{sec:conn-betw-recent}, the value of  $I(\{A_n\}_{n=1}^{N_A},\varrho)$ does not change, if the optimization in \EQ{eq:infsum} is carried out over decompositions to mixed states. 

There are efficient methods to calculate the convex roof in \EQ{eq:infsum} with semidefinite programming \cite{Toth2015Evaluating}. Calculating the minimum of \EQ{eq:infsum} for a set of constraints on the expectation values of operators $B_n$ is possible with the Hamiltonian
\begin{equation}
H_{\{\lambda_n\}_{n=1}^{N_A},\{\mu_n\}_{n=1}^{N_c}}=\sum_{n=1}^{N_A} (A_n^2 - \lambda_n A_n) - \sum_{n=1}^{N_c} \mu_n B_n,\label{eq:HAB}
\end{equation}
where $N_c$ is the number of constraints. In many cases, the lower bound on $I(\{A_n\}_{n=1}^{N_A},\varrho)$  can be obtained, analogously to \EQ{eq:Fjbound} as 
\begin{equation}
\min_{\{\lambda_n\}_{n=1}^{N_A},\{\mu_n\}_{n=1}^{N_c}:\{\ex{B_n}=b_n\}_{n=1}^{N_c}} \sum_{k=1}^{N_A} \va{A_k}_{\ket{\Psi_{\{\lambda_n\}_{n=1}^{N_A},\{\mu_n\}_{n=1}^{N_c}}}}.\label{eq:minvar}
\end{equation}
In principle, some complications might arise if the ground state of the Hamiltonian given in \EQ{eq:HAB} is degenerate or due to the fact that the minimization was restricted to pure states \footnote{If the ground state  of \EQ{eq:HAB} is degenerate then we can break the degeneracy with additional operators. Note also that \EQ{eq:minvar} works based on optimizing over pure states only. However, for some $\{\ex{B_n}=b_n\}$ constraints there might not be a corresponding pure state. Thus, when we plot $\sum_n \ex{A_n^2}$ as a function of $\ex{A_n}$  and $\ex{B_n}$ from results of the optimization over pure states, we have to construct the convex hull from below. Then, we can obtain the function $f$ giving the minimum as $(\sum_n \ex{A_n^2})_{\min}=f(\{\ex{A_n}\}_{n=1}^{N_A},\{\ex{B_n}\}_{n=1}^{N_c})$ that is valid even for mixed states. We can then compute the minimal variance based on this function rather than using \EQ{eq:minvar}.}.
\REFL{Apellaniz2017Optimal}  considers a similar problem, but uses the Legendre transform instead of Lagrange multipliers for  the case of a single $A_n$ operator. The method can straightforwardly be generalized to the case of several $A_n$ operators.

Finally, we can obtain a similar relation with a maximization  rather than a minimization over the decomposition. 

\begin{observation}
\label{obs:convevroof_multiterm2}
Let us consider a relation
\begin{equation}
\sum_{n=1}^{N_A} \va{A_n}_\varrho \le h(\varrho),
\end{equation}
which is true for pure states with some $A_n$ operators. If $h(\varrho)$ is concave in density matrices, then
\begin{equation}
R(\{A_n\}_{n=1}^{N_A},\varrho) \le h(\varrho)
\end{equation}
holds for mixed states, where we define via a concave roof the quantity 
\begin{equation}
R(\{A_n\}_{n=1}^{N_A},\varrho)= \max_{\{p_k,\ket{\psi_k}\}}\sum_k p_k \sum_{n=1}^{N_A} \va{A_n}_{\psi_k}.\label{eq:infsum2}
\end{equation}
If  $h(\varrho)$ is not concave in $\varrho,$  the inequality with a concave roof 
\begin{equation}
R(\{A_n\}_{n=1}^{N_A},\varrho) \le \max_{\{p_k,\ket{\psi_k}\}}  \sum_k p_k h(\ket{\psi_k}) \label{eq:vag22b}
\end{equation}
still holds.
\end{observation}

{\it Proof.} The proof is analogous to that of \OBS{obs:convevroof_multiterm}. $\qed$

Clearly, for a single operator
\begin{equation}
R(\{A\},\varrho)=\va{A}_{\varrho}
\end{equation}
holds. It can be shown that if we have only two operators then \cite{Leka2013Some,Petz2014,Toth2015Evaluating}
\begin{equation}
R(\{A_1,A_2\},\varrho)=\va{A_1}_{\varrho}+\va{A_2}_{\varrho}.
\end{equation}
For three observables, $R(\{A_1,A_2,A_2\}$ can be smaller than the sum of the variances
\begin{equation}
R(\{A_1,A_2,A_2\},\varrho)\le\va{A_1}_{\varrho}+\va{A_2}_{\varrho}+\va{A_3}_{\varrho}.
\end{equation}

Let us see a simple application for entanglement detection.

\begin{observation}
For separable states for $N$ spin-$j$ particles 
\begin{equation}
R(\{J_x,J_y,J_z\},\varrho)\ge Nj\label{eq:Vxyz}
\end{equation}
holds, which has been presented in \REF{Toth2015Evaluating}.  Any state violating the inequality \EQ{eq:Vxyz} is entangled.
\end{observation}

{\it Proof.} We know that for pure product states of $N$ spin-$j$ particles we have \cite{Toth2004Entanglement,Toth2007Optimal,Toth2010Generation,Vitagliano2011Spin}
\begin{equation}
\va{J_x}+\va{J_y}+\va{J_z}\ge Nj\label{eq:varJxyz}
\end{equation}
Thus, \EQ{eq:Vxyz} is true for pure product states. Since $R(\{J_x,J_y,J_z\},\varrho)$ is concave in $\varrho,$ it is also true for separable states, which are just mixtures of product states. $\qed$

The left-hand side of the relation with there variances given in \EQ{eq:varJxyz} is not smaller than the left-hand side of the criterion with $R(\{J_x,J_y,J_z\},\varrho)$ given in \EQ{eq:Vxyz}, and in some cases it is larger. Hence, the condition given in \EQ{eq:Vxyz} detects all states that are detected as entangled by \EQ{eq:varJxyz}, and it detects some further states.

\section{Metrological usefulness and entanglement conditions}

\label{sec:var_metrlogy}

In this section, we will connect the violation of uncertainty-based entanglement criteria to the metrological usefulness of the quantum state. With these findings, we address an important problem of entanglement theory:  even if entanglement is detected, it is not yet sure that the entanglement is useful for some quantum information processing task or quantum metrology \cite{Pezze2009Entanglement}. We will discuss first entanglement conditions for two bosonic modes, then entanglement criteria for two spins.

\subsection{Two-mode quantum states}

In this section, we will consider continuous variable systems. A bosonic mode can be described by the canonical $x$ and $p$ operators. For coherent states, $\ket{\alpha}$
\begin{equation}
\va{x}=\va{p} = \frac{1}{2}\label{eq:coh12}
\end{equation}
holds.  For mixtures of coherent states 
\begin{equation}
\varrho_{\rm mc}=\sum_k p_k \ketbra{\alpha_k}
\label{eq:cohmixed}
\end{equation}
we have, due to the concavity of the variance and the convexity of the QFI
\begin{equation}
\va{x}, \va{p}\ge \frac{1}{2},\quad F_Q[x,\varrho], F_Q[p,\varrho]\le 2.
\end{equation}

Let us now consider a two-mode system with the position and momentum operators $x_1, p_1, x_2, p_2.$ 

\begin{observation}
For a mixture of products of coherent states $\alpha_k^{(l)}$ of the form
\begin{equation}
\varrho_{\rm sepc}=\sum_k p_k \ketbra{\alpha_k^{(1)}} \otimes \ketbra{\alpha_k^{(2)}}
\label{eq:cohsep}
\end{equation}
the collective variances of the position and momentum are bounded from below as
\begin{equation}
\vasq{(x_1\pm x_2)}\ge 1 ;\;\;\;\; \vasq{(p_1\pm p_2)}\ge 1.
\label{eq:vaxp} 
\end{equation}
Moreover, the QFI for the same operators is bounded from above as
\begin{equation}
F_Q[\varrho,p_1\pm p_2] \le4;\;\;\;\; F_Q[\varrho,x_1\pm x_2]\le4.
\label{eq:FQxp}  
\end{equation}
Note that for such states the multi-variable Glauber-Sudarshan $P$ function is non-negative \cite{Agudelo2013Quasiprobabilities}.
\end{observation}
 
{\it Proof.} For a coherent state, for the variances of $x$ and $p$ the relation in \EQ{eq:coh12} holds.
Then, for a tensor product of two coherent states we have
\begin{equation}
\vasq{(x_1\pm x_2)}=\vasq{(p_1\pm p_2)} = 1.
\end{equation}
Since for pure states the QFI is four times the variance, for a tensor product of two coherent states we have
\begin{equation}
F_Q[\varrho,x_1\pm x_2]=F_Q[\varrho,p_1\pm p_2]= 4.
\end{equation}
Then, the statement follows from the concavity of the variance and the 
convexity of the QFI. $\qed$

Let us now consider entanglement detection in such systems with uncertainty relations. A well-known entanglement criterion is \cite{Duan2000Inseparability,Simon2000Peres-Horodecki}
\begin{equation}
 \vasq{(x_1+x_2)} + \vasq{(p_1-p_2)} \ge 2.\label{eq:Duan}
\end{equation}
If a quantum state violates \EQ{eq:Duan}, then it is entangled.

Next, let us connect the violation of \EQ{eq:Duan} to the metrological properties of the quantum state.
  
\begin{observation}
For a two-mode state, the following uncertainty relation holds
\begin{eqnarray}
&&\vasq{(x_1+x_2)} +\vasq{(p_1-p_2)} \ge \nonumber\\
&&\;\;\;\;\;\;\;\;\;\;4/F_Q[\varrho,p_1+p_2] + 4/F_Q[\varrho,x_1-x_2]  .
\label{eq:vaxpFxp_product} 
\end{eqnarray}
\end{observation}
As a consequence of \EQ{eq:vaxpFxp_product}, states violating the entanglement condition given in \EQ{eq:Duan} are metrologically more useful than states of the form given in \EQ{eq:cohsep}, i.e., bipartite states with a non-negative multi-variable Glauber-Sudarshan $P$ function.

{\it Proof.} We start from the relations
\begin{subequations}
\begin{eqnarray} 
\vasq{(x_1+x_2)} F_Q[\varrho,p_1+p_2] &\ge& 4,\label{eq:varFQ_simple1}\\
\vasq{(p_1-p_2)}  F_Q[\varrho,x_1-x_2] &\ge& 4,\label{eq:varFQ_simple2}\end{eqnarray}\label{eq:varFQ_simple}\end{subequations}which are the applications of \EQ{eq:varFQproductUncRel}. Then,  in both inequalities of \EQ{eq:varFQ_simple} we divide by the term containing the QFI. Finally, we sum the two resulting inequalities. 

Next, we will show that violating the condition given in \EQ{eq:Duan} implies metrological usefulness compared to a special class of separable states. Due to \EQ{eq:vaxpFxp_product}, the violation of the entanglement criterion given in \EQ{eq:Duan} implies the violation of one of the inequalities of \EQ{eq:FQxp}. Thus, violation of the uncertainty relation-based entanglement condition also means that the state has larger metrological usefulness than states of the type given in \EQ{eq:cohsep}. $\qed$
  
Note however that we did not prove that violating the entanglement condition given in \EQ{eq:Duan} leads to larger metrological usefulness than that of separable states in general, since even for pure product states $F_Q[\varrho,x_1\pm x_2]$ or $F_Q[\varrho,p_1\pm p_2]$ can be arbitrarily large for two bosonic modes.
  
\subsection{Spin systems}
  
Next, we will consider a system of two spins. For this case, we can show that if entanglement is detected by a well-known entanglement condition, then the state is more useful for metrology than a certain subset of separable states.
  
Let us see first a well-known entanglement conditions for two spins \cite{Toth2004Entanglement}. For separable states 
\begin{eqnarray}
\label{eq:varjxjyjz_bipartite}
&&\vasq{(J_x^{(1)}+J_x^{(2)})}+\vasq{(J_y^{(1)}+J_y^{(2)})}+\vasq{(J_z^{(1)}+J_z^{(2)})}\nonumber\\
&&\quad\quad\quad\ge j_1+j_2.
\end{eqnarray}
holds. Any state violating \EQ{eq:varjxjyjz_bipartite} is entangled. Note that this is the same condition as \EQ{eq:varJxyz} for the special case of two qudits.

Next, we need a similar relation for the QFI. For that, let us consider a special class of mixed states, a mixture of spin coherent states of a spin-$j$ particle given as
\begin{equation}
\varrho_{\rm msc}=\sum_k p_k \ketbra{s_k}.
\label{eq:spincohmix}
\end{equation}
Here, the spin-coherent states are defined similarly as in \EQ{eq:spincoherent}. It is easy to see that for such states 
\begin{equation}
 \sum_{l=x,y,z}F_Q[\varrho,j_l]\le 4j\label{eq:FQQ}
\end{equation}
holds, where the inequality is saturated for all pure spin coherent states. The maximum of the left-hand side of \EQ{eq:FQQ} for general quantum states is $4j(j+1).$ We add that for spin-coherent states 
\begin{equation}
F_Q[\varrho,j_l]\le 2j\label{eq:FQQ2}
\end{equation}
also holds for $l=x,y,z,$ where  the inequality is saturated for $\ket{+j}_k$ for $k\ne l.$ The maximum of the left-hand side of \EQ{eq:FQQ2} for general quantum states is $4j^2.$

Let us now move to bipartite systems.

\begin{observation} 
For a mixture of products of spin-coherent states $\ket{s_k^{(l)}}$ of the form
\begin{equation}
\varrho_{\rm sepsc}=\sum_k p_k \ketbra{s_k^{(1)}} \otimes \ketbra{s_k^{(2)}},
\label{eq:spincohsep}
\end{equation}
the relation with the sum of three QFI terms
\begin{eqnarray}
\label{eq:sep3F}
&& F_Q[\varrho,J_x^{(1)}\pm J_x^{(2)}]+F_Q[\varrho,J_y^{(1)}\pm J_y^{(2)}] 
+F_Q[\varrho,J_z^{(1)}\pm J_z^{(2)}]\nonumber\\
&&\quad\quad\quad \le 4(j_1+j_2)
\end{eqnarray}
holds.   
\end{observation}

{\it Proof.} This is just a generalization of the statements presented in  \REFS{Hyllus2012Fisher,Toth2012Multipartite}. For a pure product of spin-coherent states, for the left-hand side of \EQ{eq:sep3F} we have
\begin{eqnarray}
4 \left[\sum_{l=x,y,z} \va{J_l^{(1)}}+\sum_{l=x,y,z} \va{J_l^{(2)}}\right]= 4(j_1 +j_2).\nonumber\\\label{eq:varvar}
\end{eqnarray}
Due to the convexity of the QFI, the left-hand side of \EQ{eq:sep3F} cannot be larger than the right-hand side even for mixed states. $\qed$

A state violating the inequality given in \EQ{eq:sep3F} is more useful metrologically than a mixture of products of spin-coherent states, if we consider not a single metrological task, but the three tasks corresponding to the three QFI terms in \EQ{eq:sep3F}. 

Next, we will show how the violation of \EQ{eq:varjxjyjz_bipartite} implies metrological usefulness.

\begin{observation} For a bipartite quantum state 
\begin{eqnarray}
&&8 \sum_{l=x,y,z} \vasq{(J_l^{(1)}+J_l^{(2)})}+\sum_{l=x,y,z} F_Q[\varrho,J_l^{(1)}-J_l^{(2)}]\nonumber\\
  &&\quad\quad\quad \ge 12 (j_1+j_2)
  \label{eq:varJxJyJzFQxyz} 
  \end{eqnarray}
holds. Here, $J_l^{(n)}$ for $l=x,y,z$ are spin operators acting on the two subsystems, and $j_n$ are the spins of the two parties. 

As a consequence of \EQ{eq:varJxJyJzFQxyz}, states violating the entanglement condition in  \EQ{eq:varjxjyjz_bipartite} are metrologically more useful than mixtures of products of spin-coherent states given in \EQ{eq:spincohsep}, which are a subset of separable states, for the combination of the three metrological tasks corresponding to the three QFI terms in \EQ{eq:varJxJyJzFQxyz}. For the case $j_1=j_2=1/2,$ this also means that they are more useful than separable states.
\end{observation}

{\it Proof. } We start from the uncertainty relations for the two parties
\begin{equation}
\va{J_x^{(n)}}+\va{J_y^{(n)}}+\va{J_z^{(n)}} \ge j_n,
\end{equation}
where $n=1,2.$ For pure states of spin-1/2 particles, the equality holds. Then, we need the fact that
\begin{eqnarray}
&&F_Q[\varrho,J_x^{(1)}-J_x^{(2)}]/4+\vasq{(J_y^{(1)}+J_y^{(2)})}+\vasq{(J_z^{(1)}+J_z^{(2)})}\nonumber\\
&&\quad\quad\quad \ge j_1+j_2\label{eq:FQxminusVaryplusVarzplus}
\end{eqnarray}
is valid for any quantum state. This can be seen knowing that it is true for pure states, i.e.,
\begin{eqnarray}
\label{eq:varjxjyjz_bipartite2}
&&\vasq{(J_x^{(1)}-J_x^{(2)})}+\vasq{(J_y^{(1)}+J_y^{(2)})}+\vasq{(J_z^{(1)}+J_z^{(2)})}\nonumber\\
&&\quad\quad\quad \ge j_1+j_2,
\end{eqnarray}
which can be proved similarly to  \EQ{eq:varjxjyjz_bipartite}. Then, the mixed state condition follows from ideas of \SEC{sec:3varfisher}. Using \EQ{eq:FQxminusVaryplusVarzplus}, and all the inequalities obtained from it after permuting $x,$ $y$ and $z,$ and adding these inequalities, we arrive at \EQ{eq:varJxJyJzFQxyz}. 

Let us see the second part of the Observation. If \EQ{eq:varjxjyjz_bipartite} is violated, then based on \EQ{eq:varJxJyJzFQxyz}
\begin{equation}
\sum_{l=x,y,z}F_Q[\varrho,J_l^{(1)}-J_l^{(2)}] > 4(j_1+j_2), 
\end{equation}
must hold. We know that for a mixture of products of spin-coherent states the inequality given in \EQ{eq:sep3F} holds. Thus, the quantum states violating the entanglement condition given in \EQ{eq:varjxjyjz_bipartite} are more useful for metrology than  states of the form
\EQ{eq:spincohsep}.
$\qed$

Let us examine \EQ{eq:varJxJyJzFQxyz} for SU(2) singlet states.
For such states, 
\begin{equation}
\exs{(J_l^{(1)}+J_l^{(2)})^2}=0
\end{equation}
for $l=x,y,z.$ Hence, for such states the first sum in \EQ{eq:varJxJyJzFQxyz} is zero, and 
\begin{equation}
\sum_{l=x,y,z}F_Q[\varrho,J_l^{(1)}-J_l^{(2)}]\ge 12(j_1+j_2).
\end{equation}
Hence singlet states violate \EQ{eq:sep3F} with the choice of "-" for all the three terms.

Singlets are invariant under Hamiltonians of the type 
\begin{equation}
H_{0}=B_0(J_l^{(1)}+J_l^{(2)}),
\end{equation}
which describes the effect of homogeneous magnetic fields, where $B_0$ is a constant proportional to the strength of the homogeneous magnetic field. However, singlet states are sensitive to field gradients \cite{Cable2010Parameter,Urizar-Lanz2013Macroscopic,Behbood2014Generation}.

\section{Conclusions}

We studied various relations obtained from the Schr\"odinger-Robertson uncertainty after an optimization over all the possible decompositions of the density matrix is applied. Using convex roofs over decompositions,  we rederived the inequality presented in \REF{Frowis2015Tighter}, and gained insights concerning the Cram\'er-Rao bound. We also used concave roofs to obtain improvements  on the Robertson-Schr\"odinger uncertainty relation.  Finally, using similar techniques, we introduced inequalities with variances and the QFI. Similar techniques might make it possible to obtain inequalities for variances and the QFI from further inequalities for variances \cite{Dammeier2015UncertaintyNJP}.

Independently from our work, the convex-roof property of the QFI has been used to derive uncertainty relations by Chiew and Gessner \cite{Chiew2021ImprovingB}.

\acknowledgments

We thank I. Apellaniz, R. Demkowicz-Dobrza\'nski, P. Hyllus, O. G\"uhne,  M. Kleinmann, J. Ko\l ody\'nski,  J. Siewert, A. Smerzi, Sz. Szalay, R. Tr\'enyi, and G. Vitagliano  for stimulating discussions.  We acknowledge the support of the  EU (COST Action CA15220, QuantERA CEBBEC, QuantERA MENTA), the Spanish MCIU (Grant No. PCI2018-092896), the Spanish Ministry of Science, Innovation and Universities and the European Regional Development Fund FEDER through Grant No. PGC2018-101355-B-I00 (MCIU/AEI/FEDER, EU), the Basque Government (Grant No. IT986-16), and the National Research, Development and Innovation Office NKFIH (Grant No.  K124351, No. KH129601).  We thank the "Frontline" Research Excellence Programme of the NKFIH (Grant No. KKP133827). G.T. is thankful for a  Bessel Research Award from the Humboldt Foundation.

\appendix

\section{Derivation of  \EQ{eq:Heisenberg2}}
\label{App:A}

We derive   \EQ{eq:Heisenberg2} from knowing that the Robertson-Schr\"odinger inequality given in \EQ {eq:RS1} holds for all $\varrho_k$ components.  

Let us consider the inequality 
\begin{equation} \label{eq:ineq}
\left(\sum_k p_k a_k \right)\left(\sum_k p_k b_k \right)\ge\left(\sum_k p_k \sqrt{a_k b_k}\right)^2,
\end{equation}
where $a_k,b_k\ge0.$ It can be proved as follows. It can be rewritten as 
\begin{equation}
\sum_{k,l} p_k  p_l (a_k b_l + a_l b_k)  \ge \sum_{k,l} p_k  p_l 2\sqrt{a_k a_l b_l b_k}.
\end{equation}
Term by term, the left-hand side is larger or equal to the right-hand side, since $(\sqrt{a_k b_l}-\sqrt{ a_l b_k})^2\ge0.$ If additionally  
\begin{equation}
a_k b_k\ge c_k^2 \label{eq:abc}
\end{equation}
holds for all $k$ then we arrive at 
\begin{equation} \label{eq:ineq2}
\left(\sum_k p_k a_k \right)\left(\sum_k p_k b_k \right)\ge\left(\sum_k p_k \vert c_k \vert \right)^2.
\end{equation}

Note that \EQ{eq:ineq}, and hence  \EQ{eq:ineq2} can be saturated only if 
\begin{eqnarray}
a_k&=&a_l,\nonumber\\
b_k&=&b_l,
\end{eqnarray}
hold for all $k,l.$ Finally, in order to have equality in \EQ{eq:ineq2}, we also need  that \EQ{eq:abc} is saturated for all $k.$ In this case, all $c_k$ must be equal to each other.

The inequality  in \EQ{eq:Heisenberg2} can be derived from the relation in \EQ{eq:ineq2} knowing that the uncertainty relation given in \EQ {eq:RS1} holds for the $\varrho_k$ components in a decomposition given in \EQ{eq:rhodecomp}. We need to introduce $a_k=\va{A}_{\varrho_k},$ $b_k=\va{B}_{\varrho_k},$ and $c_k=\frac{1}{2}L_{\varrho_k}.$ If we use an inequality analogous to \EQ{eq:Heisenberg2} for pure-state decompositions given in \EQ{decomp} then we need $a_k=\va{A}_{\psi_k},$ $b_k=\va{B}_{\psi_k},$ and $c_k=\frac{1}{2}L_{\psi_k}.$

\section{Numerical calculation of concave roofs}
\label{sec:numerical}

In this appendix, we will discuss how to compute concave roofs numerically. Concave roofs can be computed by brute force optimization.  We will now describe a simple numerical method to find such bounds. Other method is similar to the one in \REF{Rothlisberger2009Numerical}, as it is also based on the purification of the mixed state. The statements also hold for concave roofs, after trivial changes. 

In order to obtain concave roofs, we have to carry out a numerical optimization over all decompositions of the density matrix. First let us consider decompositions to pure states given in \EQ{decomp}. Let us define the purification of $\varrho$ \cite{Hughston1993AComplete}
\begin{equation}
\ket{\Psi_p}=\sum_k \sqrt{p_k} \ket{\psi_k}_S \otimes \ket{k}_A,
\end{equation}
where $S$ denotes the system, $A$ is the ancilla and for this state, 
\begin{equation}
\trace_A(\ketbra{\Psi_p})=\varrho
\end{equation}
holds. One of the purifications that is easy to write is the one based on the eigendecomposition of the density matrix, and for that we need an ancilla that has the same size as the system. For other purifications, we might need an ancilla larger than the system. The dimension of the ancilla equals the number of pure subensembles we consider.

Since all purifications can be obtained from each other by a unitary acting on the ancilla, we arrive at the following. For any quantity $Q(\sigma),$ which is a function of a mixed state $\sigma$ we can write the concave roof as an optimum over the decompositions as 
\begin{eqnarray}
&&\max_{\{p_k,\ket{\psi_k}\}} \sum_k p_k Q(\ketbra{\psi_k})\nonumber\\
&&\quad\quad=\max_{U_A} \sum_k \langle v_k \vert v_k \rangle Q(\ketbra{v_k}/\langle v_k \vert v_k \rangle),\quad\label{eq:supremum}
\end{eqnarray}
where the maximization is over unitaries acting on the ancilla and we defined the unnormalized vectors as
\begin{equation}
\ket{v_k}={\bra{k}}_A U_A \ket{\Psi_p}.
\end{equation}
Note that one can show that
\begin{equation}
\varrho=\sum_k \ketbra{v_k}.
\end{equation}

These ideas can be extended to mixed-state decompositions given in  \EQ{eq:rhodecomp} as follows. Similarly to \SEC{sec:RS}, we consider not only pure-state decompositions, but also mixed-state decompositions in which the mixed components are mixtures of some of the $\vert v_k\rangle.$ 
We can extend this method to optimize over all mixed-state decompositions as follows
\begin{eqnarray}
&&\max_{\{K_l\}} \max_{\{p_l,\rho_l\}} \sum_l p_l Q(\varrho_l)\nonumber\\
&&\quad\quad=\max_{\{K_l\}} \max_{U_A} \sum_l \trace(\sigma_l) Q[\sigma_l/\trace(\sigma_l)],\quad\label{eq:supremum2}
\end{eqnarray}
where unnormalized states are 
\begin{equation}
\sigma_l=\sum_{k\in K_l}  {\bra{k}}_A U_A \ket{\Psi_p}\bra{\Psi_p} U_A^\dagger\ket{k}_A.
\end{equation}
The probabilities and the normalized states of the decomposition are given as 
\begin{equation}
p_k={\rm Tr}(\sigma_k),\quad\quad\quad\varrho_k=\sigma_k/p_k.
\end{equation}
Here the basis states are  distributed into sets $K_l.$ For instance, $K_1=1, K_2=2,$ and $K_3=3$ corresponds to looking for a pure-state decomposition.
$K_1=\{1,2\},$ and $K_2=3$ corresponds to looking for a mixture of a rank-2 mixed state and a pure state. In \FIG{fig:rs}(b), the results are shown for using the method above where both the system and the ancilla have a dimension $d=3.$

Looking for the unitary that leads to the maximum can be done  with a multivariable search. We developed a simple algorithm based on a random search, and improving the best random guess by small local changes. The local changes are also random and they are accepted if they increase the quantity to be maximized. A computer program based on such an algorithm is incorporated in the newest version of the QUBIT4MATLAB package \cite{*[{The name of the functions relevant for this publication are {\tt concroof.m} and {\tt example\_concroof.m}, respectively. The program package is available at MATLAB CENTRAL at {\tt http://www.mathworks.com/matlabcentral/}. The 3.0 version of the package is described in }] [{.}] Toth2008QUBIT4MATLAB}. Such random optimization has already been used to look for the maximum of an operator expectation value for separable states in the same program package.

\REFL{Toth2015Evaluating} presents a method that provides good upper bounds of concave roofs based on semidefinite programming, but it works only for small systems of a couple of qubits. The result of this procedure is larger or equal to the true bound and thus can be used to evaluate whether the bound found with the brute force search is optimal.

Calculating the convex roof is similar, only the maximization has to be replaced by minimization in \EQ{eq:supremum}.

\bibliography{Bibliography2}
\end{document}